\documentclass[aip,jcp,reprint,groupedaddress,floatfix]{revtex4-1}
\pdfoutput=1

\usepackage{siunitx} % Provide correct typesetting for SI units

\usepackage{amsmath} % Equation typesetting
\usepackage{amssymb}
\usepackage{mathrsfs}
\usepackage{mathtools}
\usepackage{textcomp}
\usepackage{stmaryrd}
\usepackage{booktabs}
\usepackage{esint}

\usepackage[shortlabels]{enumitem}
\usepackage{graphicx} % Image handler

\usepackage{multirow}
\usepackage{booktabs} % Horizontal rules in tables
\usepackage{hyperref} % For hyperlinks in the PDF

\newcommand*\chem[1]{\ensuremath{\mathrm{#1}}}

\DeclareSIUnit{\molar}{M}

\AtBeginDocument
{
  \heavyrulewidth=.08em
  \lightrulewidth=.05em
  \cmidrulewidth=.03em
  \belowrulesep=.65ex
  \belowbottomsep=0pt
  \aboverulesep=.4ex
  \abovetopsep=0pt
  \cmidrulesep=\doublerulesep
  \cmidrulekern=.5em
  \defaultaddspace=.5em
}

\begin{document}
\title{Incorporating particle flexibility in a density functional description of nematics and cholesterics}
\date{\today}
\author{Maxime M.C. Tortora}
\author{Jonathan P.K. Doye}
\affiliation{Physical and Theoretical Chemistry Laboratory, Department of Chemistry, University of Oxford, South Parks Road, Oxford OX1 3QZ, United Kingdom}

\begin{abstract}
We describe a general implementation of the Fynewever-Yethiraj density functional theory (DFT) for the investigation of nematic and cholesteric self-assembly in arbitrary solutions of semi-flexible polymers. The basic assumptions of the theory are discussed in the context of other generalised Onsager descriptions for flexible polyatomic systems. The location of the isotropic-to-nematic phase transition is found to be in good agreement with molecular simulations for elongated chains up to relatively high polymer flexibilities, although the predictions of the theory in the nematic regime lead to gradual underestimations of order parameters with decreasing particle stiffness. This shortcoming is attributed to increasing overestimations of the molecular conformational entropy in higher-density phases, which may not be easily addressed in the formalism of DFT for realistic particle models. Practical consequences of these limitations are illustrated through the application of DFT to systems of near-persistence-length DNA duplexes, whose cholesteric behaviour is found to be strongly contingent on their detailed accessible conformational space in concentrated solutions. 

 \vspace{2mm}
 
 \begin{center}
  \includegraphics[width=25pc]{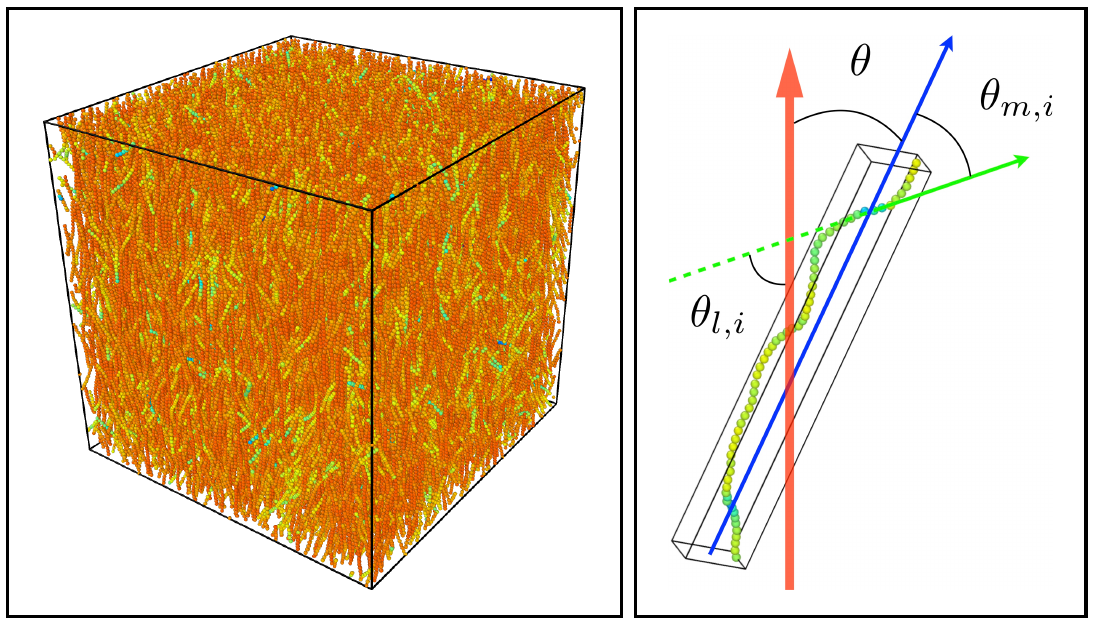}
\end{center}

\end{abstract}


\maketitle

\section{Introduction} \label{sec:Introduction}
The self-organisation of polymer solutions into partially-ordered mesophases is a phenomenon of considerable biological relevance, whose occurrences span the formation of phospholipid bilayers for the assembly of cell plasma membranes\cite{Goet98}, the arrangements of F-actin filaments in the cytoplasm\cite{Copp92} and of haemoglobin chains in sickle-cell anaemic blood\cite{Harr50}. This natural ubiquity, combined with a sustained industrial interest in the practical applications of polymeric liquid crystals\cite{Cife82,Coll92,Dona06}, has spawned a wealth of fundamental and applied investigations of their phase behaviour in a variety of contexts\cite{Figu05,Lage12}.
\par
The first successful mathematical account of emerging order in polymer solutions stems from the seminal work of Onsager\cite{Onsa49}, who inferred from first-principle arguments the existence of a concentration-driven transition from a liquid-like isotropic state to an orientationally-ordered nematic state in dispersions of anisotropic colloids. In this framework, which preludes the classical density functional theory (DFT) of non-uniform fluids, the onset of nematic organisation simply results from the competition between orientational entropy and excluded-volume interactions, for which exact explicit expressions may be obtained in the limit of rigid rod-like particles with infinite particle aspect ratios. 
\par
Real polymers, however, are generally characterised by a finite flexibility, which may lead to substantial deviations from the slender linear conformations considered by Onsager, and which needs to be accurately incorporated in any reliable statistical-mechanical description of their self-assembly. Considerable efforts have thus been devoted to the generalisation of the Onsager theory to systems of limited stiffness based on various coarse-grained polymer representations and simplified derivations of the corresponding conformational free energy, which provides an additional contribution to the Onsager thermodynamical treatment of their isotropic-to-nematic (I/N) transition\cite{Khok78,Khok81,Khok82,Khok84,Odij86,Hent90,Sato90,Dupr91,Chen93,vanD94,Fyne98,Jaff01,Dipl04,Denn11,Zhan15}. 
\par
The approximate nature of such approaches therefore generally requires their careful preliminary assessment against the results of molecular simulations in order to probe the quantitative effects of their potential shortcomings, independently from the inevitable ambiguities of direct comparisons with experimental results\cite{Egor16-1}. However, most current formulations of DFT generally rely on a number of system-specific analytical assumptions and numerical approximations, which do not allow for the straightforward application of theory and simulations to identical particle models. Furthermore, theoretical investigations of the nematic behaviour of semi-flexible chains have so far almost exclusively focused on their uniaxial organisation, while the liquid-crystal phases formed by many biopolymers are commonly observed to rather display a chiral \textit{cholesteric} arrangement, reflecting the detailed symmetries of their local molecular structures\cite{deGe93,Mito17}.
\par
A classical extension of the Onsager theory to the cholesteric assembly of rigid particles was introduced by Straley\cite{Stra76}, who proposed to account for the macroscopic breaking of mirror symmetry in the weak chirality limit based on a microscopic statistical theory of the Frank elastic constants\cite{Stra73-1}. This perturbative treatment provides a convenient common theoretical formalism in which uniaxial nematic phases are simply assimilated to ``degenerate'' cholesteric phases with infinite helical pitch, and in which the reliable determination of equilibrium cholesteric structure is contingent on the accurate account of local uniaxial order\cite{Stra76}.
\par
However, in the case of semi-flexible particles, the difficulty of representing the dependence of the conformational free energy on their detailed chiral structure\cite{Osip94} has limited most theoretical investigations of cholesteric ordering to polymer solutions in the coil limit\cite{Odij87,Pelc96} --- while many experimental cholesteric systems are appreciably stiffer\cite{Sato98}. Furthermore, the complex virial-type coefficients underpinning macroscopic twist in Straley's approach\cite{Stra76} have only been evaluated analytically for a handful of simplified model systems\cite{Osip85,Osip94,Pelc96,Varg06,Wens09,Wens11,Wens14}, and are often used as semi-empirical adjustable parameters when comparing theoretical predictions to experimental cholesteric pitch measurements\cite{Sato98}. Therefore, a general and reliable framework to investigate the link between phase symmetry breaking and molecular chirality in solutions of semi-flexible polymers is still largely lacking.
\par
Building on previous efforts\cite{Tort17-1,Tort17-2}, we here propose an efficient implementation of DFT that may be conveniently applied to a wide variety of flexible particle models, taking into account the full details of their microscopic Hamiltonian description. In this hybrid approach, conformational statistics may be introduced through the use of molecular trajectories obtained from direct simulations, thus allowing for a rigorous comparison of theoretical predictions with simulation results. A particular advantage of the method is that it is easily generalisable to Straley's description of the cholesteric phase, and is suitable for the treatment of arbitrary polymer systems with various degrees of molecular complexity.
\par
The structure of the paper is organised as follows. We first summarise in Sec.~\ref{sec:DFT} the main attempts at the inclusion of particle flexibility in the context of DFT, with a particular emphasis on the physical assumptions underlying the different theories, and outline the details of our chosen approach and numerical implementation. We then dedicate Sec.~\ref{sec:nematic} to the extensive comparison of our results with molecular simulations of the nematic assembly of coarse-grained semi-flexible chains, and present in Sec.~\ref{sec:dna} the application of our method to the cholesteric organisation of near-persistence-length DNA duplexes. Finally, we recapitulate in Sec.~\ref{sec:conclusion} the main conclusions of these analyses, and highlight some potential directions for future research.

\section{Density functional theory for semi-flexible polymers} \label{sec:DFT}
In the framework of classical DFT, the self-organisation of polyatomic molecules in the absence of external fields is driven by the competition between two coupled contributions, generally referred to as the \textit{ideal} and \textit{excess} free energies $\mathscr{F}_{\rm id}$ and $\mathscr{F}_{\rm exc}$, and often defined as the respective intra- and inter-molecular components of their total Helmholtz free energy $\mathscr{F}$\cite{Chan86-1,Chan86-2},
\begin{equation}
  \label{eq:free}
  \mathscr{F} = \mathscr{F}_{\rm id} + \mathscr{F}_{\rm exc}.
\end{equation} 
In the so-called Onsager limit of infinitely-stiff particles with high aspect ratios assembling into uniform uniaxial nematic phases, exact analytical expressions for $\mathscr{F}_{\rm id}$ and $\mathscr{F}_{\rm exc}$ may be derived in the functional form\cite{Onsa49}
\begin{align}
  \label{eq:ideal}
  \frac{\beta\mathscr{F}_{\rm id}[\psi]}{V} &= 4\pi^2 \rho \int_0^\pi d\theta \times \sin \theta \psi(\cos\theta) \nonumber  \\ &\qquad\times \Big\{\log [\rho \psi(\cos\theta)]-1 \Big\}, \\
  \label{eq:excess}
  \frac{\beta\mathscr{F}_{\rm exc}[\psi]}{V} &= -\frac{\rho^2}{2} \int_V d\mathbf{r}_{12} \oiint d\mathcal{R}_1 d\mathcal{R}_2 \times f(\mathbf{r}_{12}, \mathcal{R}_1, \mathcal{R}_2) \nonumber \\&\qquad \times \psi(\cos\theta_1) \psi(\cos\theta_2),
\end{align}
where $\beta\equiv 1/k_bT$ is the inverse temperature, $\rho$ the uniform number density and $\psi(\cos\theta)\equiv \psi(\mathbf{u}\cdot \mathbf{n})$ the orientation distribution function (ODF) describing the degree of alignment of the molecular long axes $\mathbf{u}$ about the uniform director $\mathbf{n}$. In this case, the excess free energy Eq.~\eqref{eq:excess} may fully account for inter-particle interactions at the second-virial level through the Mayer $f$-function\cite{Maye40},
\begin{equation}
  \label{eq:Mayer}
  f(\mathbf{r}_{12}, \mathcal{R}_1, \mathcal{R}_2) = \exp\Big\{-\beta U(\mathbf{r}_{12}, \mathcal{R}_1, \mathcal{R}_2)\Big\} - 1,
\end{equation}
with $U(\mathbf{r}_{12}, \mathcal{R}_1, \mathcal{R}_2)$ the interaction energy of two particles with respective orientations $\mathcal{R}_1$, $\mathcal{R}_2$ and relative centre-of-mass separation $\mathbf{r}_{12}$, while the ideal free energy Eq.~\eqref{eq:ideal} simply reduces to that of an ideal monatomic gas of anisotropic particles with ODF $\psi$. The equilibrium ODF $\psi_{\rm eq}$ is then obtained by functional minimisation of the corresponding free energy $\mathscr{F}[\psi]$ at fixed density $\rho$,
\begin{equation}
  \label{eq:funcmin}
   \frac{\delta \mathscr{F}}{\delta \psi(\cos\theta)}\biggr|_{\psi_{\rm eq}} = \rho V k_b T \times 4\pi^2 \lambda \sin \theta,
\end{equation}
with $\lambda$ a Lagrange multiplier ensuring the proper normalisation of the ODF. Eqs.~\eqref{eq:free}--\eqref{eq:funcmin} then immediately lead to the well-known self-consistent equation for $\psi_{\rm eq}$,
\begin{multline}
  \label{eq:selfc}
  \psi_{\rm eq}(\cos\theta) = \exp\bigg[\lambda-\frac{\rho}{4\pi^2} \int_0^\pi d\theta' \times \sin\theta' E(\theta, \theta') \\ \times \psi_{\rm eq}(\cos\theta') \bigg],
\end{multline}
with $E$ the orientation-dependent generalised excluded volume,
\begin{multline}
  \label{eq:exc}
  E(\theta_1, \theta_2) = -\int_V d\mathbf{r}_{12}  \iint_0^{2\pi} d\alpha_1 d\alpha_2 \iint_0^{2\pi} d\phi_1 d\phi_2 \\ \times f(\mathbf{r}_{12}, \mathcal{R}_1, \mathcal{R}_2),
\end{multline}
where $\theta_i$, $\phi_i$ and $\alpha_i$ denote the 3 Euler angles characterising the orientation $\mathcal{R}_i$ of particle $i$.
Under these assumptions, the equilibrium structure and thermodynamics of the phase are thus fully determined by Eqs.~\eqref{eq:free}--\eqref{eq:exc}, which may be readily solved by numerical means for a variety of particle models\cite{Tort17-1, Tort17-2}.
\par
However, in the case of particles with finite stiffness, the determination of $\mathscr{F}_{\rm id}$ requires the accurate computation of their conformational free energy and its potential dependence on the surrounding medium, which is analytically intractable for all but the simplest model systems\cite{Flor56-1,Flor56-2,Yethi02}. Furthermore, the density function $\psi$ describing the orientational properties of such molecules generally depends on their detailed internal degrees of freedom, rather than their sole long axes $\mathbf{u}$, and renders the derivation of their free energy functional $\mathscr{F}[\psi]$ a complex many-body problem even at the second-virial level. Therefore, several routes have been proposed to construct approximate expressions for $\mathscr{F}_{\rm id}$ and $\mathscr{F}_{\rm exc}$ in the case of semi-flexible particles, the main features of which we briefly review in the following.

\subsection{The Khokhlov-Semenov (KS) approach} \label{subsec:DFT_KS}

The earliest analytical attempts can be traced back to the seminal work of Flory\cite{Flor78}, based on a discretised lattice representation of polymer solutions with limited degrees of freedom. However, the many approximations underlying such a basic description have largely restricted its applicability to purely qualitative studies in the context of the I/N phase transition\cite{Flor84}. A more sophisticated treatment was subsequently introduced by Khokhlov and Semenov\cite{Khok78,Khok81,Khok82}, and further developed by several authors\cite{Odij86,Hent90,Sato90,Dupr91,Chen93}, who suggested to combine various extensions of the original Onsager excess free energy Eq.~\eqref{eq:excess} in the limit of hard rigid cylinders with a mean-field description of several flexible chain models\cite{Khok84}. In this framework, internal particle mechanics thus enter the nematic free energy Eq.~\eqref{eq:free} purely through the inclusion of a flexibility-dependent orientational entropy term in $\mathscr{F}_{\rm id}$. The resulting expression for $\mathscr{F}$ in the case of continuous worm-like chains may then be minimised through a self-consistent numerical procedure to determine the contour-dependent equilibrium ODF $\psi_{\rm eq} (s, \mathbf{t})$, quantifying the probability of finding a curvilinear segment $s\in[0,l_c]$ of the chains with fixed contour length $l_c$ pointing in direction $\mathbf{t}$\cite{Chen93}.
\par
Despite its theoretical elegance and reported successes in the description of the uniaxial nematic properties of fairly-stiff and elongated experimental systems\cite{Vroe92}, a significant shortcoming of this approach lies in its strongly coarse-grained nature. Indeed, the KS theory and its extensions effectively reduce the dependence of the phase behaviour of polymers with a given persistence length $l_p$ to a handful of simplified flexibility mechanisms\cite{Khok85} combined with the two ratios $l_c/l_p$ and $l_p/d_{\rm eff}$, with $d_{\rm eff}$ a somewhat heuristic effective chain diameter subsuming their relevant molecular features\cite{Sato96}. Its generalisation to the treatment of cholesteric order is therefore not straightforward, as the accurate description of phase chirality requires a more detailed account of local microscopic structure\cite{Osip85,Varg06,Wens09,Wens11,Wens14}, and has to our knowledge yet to be convincingly carried out beyond the limiting case of long polymer coils\cite{Odij87,Osip94,Pelc96}, for which $d_{\rm eff} \ll l_p \ll l_c$.

\subsection{The Fynewever-Yethiraj (FY) approach} \label{subsec:DFT_FY}

\begin{figure}[]
  \includegraphics[width=0.6\columnwidth]{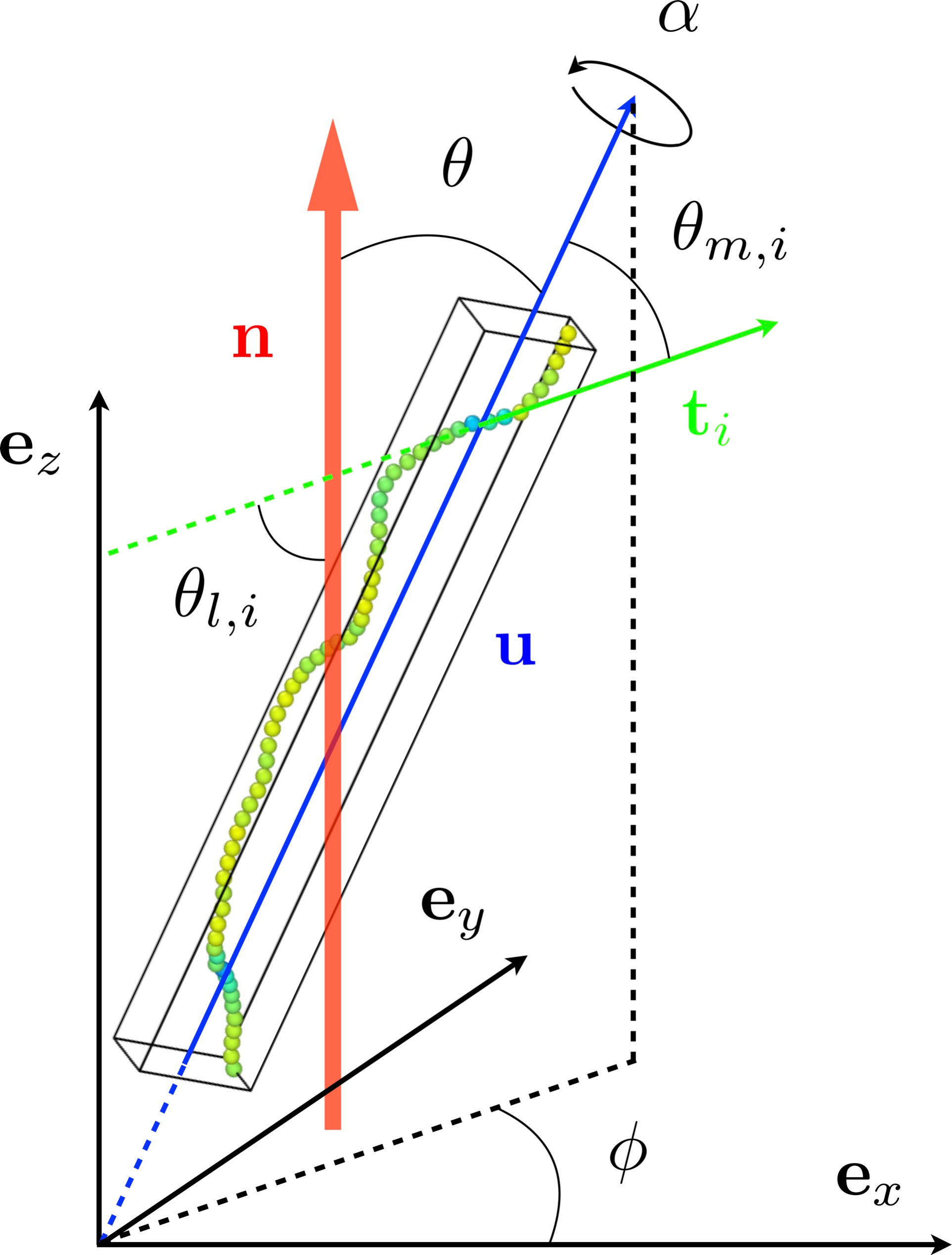}
  \caption{\label{fig1}Particle long axis $\mathbf{u}$, local bond orientations $\mathbf{t}_{i}$ and nematic director $\mathbf{n}$ for an arbitrary conformation of the coarse-grained persistent chain model described in Ref.~\onlinecite{Egor16-1}. The fixed laboratory frame $\mathcal{R}_{\rm lab} \equiv\big [\mathbf{e}_x \; \mathbf{e}_y\; \mathbf{e}_z\big]$ is defined such that $\mathbf{n}\equiv\mathbf{e}_z$. The black bounding box denotes the eigenvectors of the molecular gyration tensor $\mathcal{G}$ as determined by the numerical procedure of Sec.~\ref{subsec:DFT_procedure}, and defines the molecular frame $\mathcal{R}$.}  
\end{figure}

A somewhat orthogonal treatment was proposed by Fynewever and Yethiraj\cite{Fyne98}, who conversely suggested to account for the effects of flexibility directly through the excess free energy contribution $\mathscr{F}_{\rm exc}$, while retaining the Onsager expression for $\mathscr{F}_{\rm id}$ (Eq.~\eqref{eq:ideal}). In this context, the virial integral Eq.~\eqref{eq:excess} is instead averaged over a representative ensemble $\Omega$ of particle conformations as generated by molecular simulations, while the orientational entropy is simply determined by the chain-averaged ODF $\psi(\mathbf{u}\cdot \mathbf{n})$ describing the molecular ordering of the polymer long axes $\mathbf{u}$, as depicted in Fig.~\ref{fig1}. The corresponding excess free energy for arbitrary flexible particles may then be written in the form
\begin{align}
  \label{eq:exc_ave}
  \frac{\beta\mathscr{F}_{\rm exc}[\psi]}{V} &= G(\eta)\frac{\rho^2}{2} \iint_0^\pi d\theta_1 d\theta_2 \times \sin\theta_1\sin\theta_2 \nonumber \\ &\qquad \times \psi(\cos\theta_1) \psi(\cos\theta_2) \times \big\langle E(\theta_1,\theta_2) \big\rangle_\Omega,
\end{align}
where $\big\langle E(\theta_1,\theta_2) \big\rangle_\Omega$ denotes the ensemble average of the generalised excluded volume Eq.~\eqref{eq:exc} over $\Omega$, as discussed in the next paragraphs. Note that following Ref.~\onlinecite{Fyne98}, we have introduced in Eq.~\eqref{eq:exc_ave} the Parsons-Lee (PL) prefactor\cite{Pars79,Lee87},
\begin{equation}
  \label{eq:pl}
  G(\eta) =  \frac{1-3\eta/4}{(1-\eta)^2},
\end{equation}
with $\eta\equiv \rho v_0$ the volume fraction of the particles with molecular volume $v_0$, which provides for a simplified account of the higher-order virial corrections necessary to extend the Onsager description to mesogens with finite aspect ratios. The equilibrium properties of the resulting uniaxial phase may then be determined by solving an extended form of the Onsager self-consistent Eq.~\eqref{eq:selfc},
\begin{multline}
  \label{eq:selfc_ave}
  \psi_{\rm eq}(\cos\theta) = \exp\bigg[\lambda -G(\eta)\frac{\rho}{4\pi^2} \int_0^\pi d\theta' \times \sin\theta' \\ \times \big\langle E(\theta, \theta')\big\rangle_\Omega \psi_{\rm eq}(\cos\theta') \bigg].
\end{multline}

In this framework, the inclusion of intra-molecular free energy contributions into Eq.~\eqref{eq:free} is therefore implicitly relegated to the numerical ensemble-averaged kernel $\big\langle E(\theta_1, \theta_2) \big\rangle_\Omega$, which circumvents the analytical difficulties involved in the explicit functional derivation of $\mathscr{F}$ for non-uniform, flexible polyatomic systems\cite{Chan86-1}. Its results have been found to compare favourably with direct numerical simulations of various athermal solutions of semi-flexible chains, with quantitative agreement being reported in a number of cases\cite{Fyne98,Jian07,vanW12,vanW13}. Furthermore, its relatively simple mathematical formulation is largely independent of the specific features of the molecular system studied, and allows for its straightforward application to a wide range of mesogen models.
\par
This hybrid approach, however, is quite numerically expensive, and suffers from two inherent drawbacks. The first lies in its coarse-grained representation of orientational order at the level of the molecular long axes $\mathbf{u}$, as opposed to the more detailed description of local inter-chain correlations between neighbouring monomers through a contour-dependant ODF of the KS-based theories. The second is slightly more subtle, and stems from the tacit assumption in the derivation of Eq.~\eqref{eq:selfc_ave} from Eqs.~\eqref{eq:ideal}, \eqref{eq:funcmin} and \eqref{eq:exc_ave} that the ensemble average $\big \langle \cdot \big\rangle_\Omega$ is independent of the molecular ODF $\psi$,
\begin{equation*}
  \frac{\delta \big \langle E\big\rangle_\Omega}{\delta \psi} = 0,
\end{equation*}
which amounts to neglecting the influence of the variations of the surrounding nematic field on the accessible particle conformational space $\Omega$. In the original FY paper, $\Omega$ was therefore chosen to describe the conformations of a single chain fluctuating in free space --- an assumption we refer to as the \textit{unconfined-chain approximation} (UCA) in the rest of the paper.
\par
In order to shed some light on the respective effects of these approximations, and their potential consequences on the capability of DFT to describe realistic polymeric systems, we now outline a generic practical implementation of the FY theory applicable to arbitrary flexible particle models.

\subsection{Numerical procedure} \label{subsec:DFT_procedure}
We have recently introduced a highly-efficient numerical framework based on DFT to work out the emergent properties of uniaxial nematic and cholesteric phases from the structure of their constituent mesogens in the case of arbitrary rigid particle models\cite{Tort17-1, Tort17-2}. Its generalisation to the FY treatment of molecular flexibility may be readily performed as follows.
\begin{enumerate}
  \item Given an arbitrary polyatomic particle model described by a potential $U_{\rm tot}$ comprising both intra- and extra-molecular interaction forces,
  \begin{equation*}
    U_{\rm tot}=U_{\rm int} + U_{\rm ext},
  \end{equation*}
  we first construct an ensemble $\Omega=\Omega[U_{\rm int}]$ of uncorrelated particle conformations by means of a single-chain molecular dynamics (MD) or Monte-Carlo (MC) simulation in the canonical ensemble at fixed temperature $T$.
  \item We then compute the ensemble-averaged excluded volume integral $\big\langle E(\theta_1, \theta_2) \big\rangle_{\Omega}$ by MC sampling, discarding the intra-molecular components of $U_{\rm tot}$. In the context of Eqs.~\eqref{eq:Mayer} and \eqref{eq:exc}, a MC step amounts to drawing a random pair of conformations from $\Omega$, applying random rigid-body rotations $\mathcal{R}_1$, $\mathcal{R}_2$ and relative translation $\mathbf{r}_{12}$ to each respective conformation, and computing the interaction energy $U_{\rm ext}$ of the obtained configuration --- binning the result for $E$ using a discrete grid of polar angles $\theta_i$. 
  \item We finally plug the kernel $\big\langle E \big\rangle_{\Omega}$ into Eq.~\eqref{eq:selfc_ave}, which may be solved iteratively\cite{Herz84} to determine the equilibrium ODF $\psi_{\rm eq}$ at given $\rho$ and $T$.
\end{enumerate}
\par
In the case where the equilibrium phase of the chosen particle model is expected to be cholesteric, rather than uniaxially-symmetric, one may further complement these three steps with the computation of the ensemble-averaged Straley coefficients $\big\langle \kappa_{01} \big\rangle_{\Omega}$, $\big\langle \kappa_{11} \big\rangle_{\Omega}$, defined as\cite{Stra76}
\begin{multline}
  \label{eq:Frank}
  \beta\kappa_{ij} = -\frac{\rho^2}{2} \int_V d\mathbf{r}_{12} \oiint d\mathcal{R}_1 d\mathcal{R}_2 \times f(\mathbf{r}_{12}, \mathcal{R}_1, \mathcal{R}_2) \\ \times \psi_{\rm eq}^{(i)}(u_1^{(z)})\Big(-u_1^{(y)} r_{12}^{(x)}\Big)^i \times \psi_{\rm eq}^{(j)}(u_2^{(z)})\Big(u_2^{(y)} r_{12}^{(x)}\Big)^j,
\end{multline}
with $\mathbf{e}_x$ the cholesteric helical axis, $v^{(k)} \equiv \mathbf{v} \cdot \mathbf{e}_k$ for any vector $\mathbf{v}$ and $\psi_{\rm eq}^{(i)}$ the $i$-th derivative of $\psi_{\rm eq}$. Under the assumptions of the FY theory, the quantities $\big\langle \kappa_{ij} \big\rangle_{\Omega}$ may be conveniently computed following the same method as  $\big\langle E \big\rangle_{\Omega}$, and the equilibrium pitch $\mathcal{P}_{\rm eq}$ of the phase in the limit of weak cholesteric twist is finally given by\cite{Tort17-2}
\begin{equation}
  \label{eq:pitch}
  \mathcal{P}_{\rm eq}(\rho, T) =  2\pi \times \frac{\big\langle\kappa_{11}\big\rangle_\Omega}{\big\langle\kappa_{01}\big\rangle_\Omega}.
\end{equation}
It should be noted that in this perturbative framework, the structure of the cholesteric phase is assumed to be locally indistinguishable from that of the nematic phase, so that all liquid-crystalline properties besides the equilibrium pitch may be identically computed in both cases following the above procedure.
\par
An important prerequisite however lies in the determination of the long axes $\mathbf{u}$ of the different molecular conformations, with respect to which we choose to quantify the orientational nematic order. Several definitions for such axes have been proposed in the literature in the case of arbitrary-shaped particles, based on the spectral decomposition of the molecular polarisability\cite{Wils96, Sonn03}, inertia\cite{Wils96, Komo94} and gyration tensors\cite{Solc71, Theo85}. We here elect to make use of the latter definition, as we feel the purely-geometric character of the gyration tensor $\mathcal{G}$ to be most consistent with the framework of our equilibrium DFT description, for which molecular electronic and inertial properties are largely irrelevant. 
\par
The general expression of $\mathcal{G}$ for an arbitrary polymer conformation defined by the coordinate matrix $\mathcal{M} \equiv \{\mathbf{r}_i\}_{i \in \llbracket 1, N \rrbracket}$ of its $N$ constituent monomers reads as
\begin{equation}
  \label{eq:gyration}
  \mathcal{G}_{mn} \equiv \frac{1}{N} \sum_{i=1}^N r_i^{(m)} r_i^{(n)} = \frac{1}{N} \big(\mathcal{M} \cdot \mathcal{M}^{\sf T}\big)_{mn}
\end{equation}
for $(m,n) \in \{x,y,z\}^2$, with $\mathbf{r}_i$ the position vector of the $i$-th monomer and $\mathcal{M}^{\sf T}$ the matrix transpose of $\mathcal{M}$, assuming the origin of the frame to be such that
\begin{equation*}
  \sum_{i=1}^{N} \mathbf{r}_i = \mathbf{0}.
\end{equation*}
The particle long axis $\mathbf{u}$ is then defined as the eigenvector associated with the largest eigenvalue of $\mathcal{G}$, and may be efficiently computed from Eq.~\eqref{eq:gyration} by principal component analysis (PCA)\cite{Tort17-2} through the singular-value decomposition of $\mathcal{M}$. In the case where all the constituent monomers bear identical masses, it is easy to show that $\mathbf{u}$ also corresponds to the eigenvector associated with the smallest eigenvalue of the polymer inertia tensor, and one recovers the definition of the molecular axis employed in the original FY implementation\cite{Fyne98}.
\par
The use of the UCA at step (1) combined with optimised virial integration methods\cite{Tort17-2} for the evaluation of Eqs.~\eqref{eq:exc} and \eqref{eq:Frank} enables us to greatly reduce the overall computational expense of the procedure, and thus allows us to tackle flexible mesogen models of arbitrary complexity. We now dedicate the next two sections to the assessment of its results in the case of achiral and chiral systems with various levels of molecular resolution.

\section{The nematic behaviour of coarse-grained polymer chains} \label{sec:nematic}
A particularly suitable generic model system for such a quantitative investigation may be provided by the so-called Kremer-Grest (KG) bead-spring chain\cite{Gres86, Krem90}, which constitutes a practical discretised realisation of the original worm-like chain model\cite{Krat49-1,Krat49-2} including the effects of both excluded volume and bond flexibility. The bulk properties of their I/N transition were recently investigated in considerable detail by Egorov \textit{et al.}\cite{Egor16-1}, who combined extensive MD simulations with a simplified implementation of the FY DFT approach to perform a comprehensive mapping of their phase diagram for a wide range of contour lengths, persistence lengths and particle concentrations. The extensive comparison of simulation results with existing theoretical predictions further confirmed that the FY approach consistently outperforms other formulations of DFT in terms of overall agreement with the MD data, despite increasing discrepancies being reported for all theories in the case of highly-flexible chains.
\par
In that study, as in previous applications of the FY theory\cite{Fyne98, Jian07}, much emphasis was put on the effects of the PL approximation on the accuracy of the DFT predictions, and several forms of the prefactor Eq.~\eqref{eq:pl} were investigated --- though no single expression was conclusively found to yield the best results systematically. Such discussions are undoubtedly highly relevant, especially in the context of the reported shortcomings of the PL rescaling for systems of non-convex particles\cite{Tort17-1}. However, we here wish to primarily address the much-less documented effects of the coarse-grained description of orientational order and of the UCA, as discussed in Sec.~\ref{subsec:DFT_FY}, and thus restrict our focus to the regime of fairly-stiff chains with high aspect ratios, for which the PL approximation is expected to work reasonably well, throughout the rest of this section.
\par
In the implementation of the KG model proposed in Ref.~\onlinecite{Egor16-1}, polymers are represented as chains of beads interacting through a repulsive Weeks-Chandler-Andersen (WCA) potential,
\begin{equation*}
  u^{\rm WCA}_{ij}(r_{ij}) = 
  \begin{dcases}
    4\epsilon \Bigg [ \bigg(\frac{\sigma}{r_{ij}} \bigg)^{12} - \bigg( \frac{\sigma}{r_{ij}}\bigg)^6 + \frac{1}{4} \Bigg] & \text{if } r_{ij} < r_c \\
    0 &\text{if } r_{ij} \geq r_c
  \end{dcases},
\end{equation*}
where $\sigma$ and $\epsilon$ define the respective model units of length and energy, $r_c\equiv 2^{1/6} \sigma$ and $r_{ij}$ denotes the separation distance between any two distinct monomers $i$ and $j$. Particle flexibility is then governed by a simplified bond-bending potential of the form\cite{Honn90}
\begin{equation*}
  u^{\rm bend}_i(\theta_i) = \epsilon_b (1-\cos\theta_i),
\end{equation*}
with $\epsilon_b$ a bending stiffness parameter and $\theta_i$ the angle formed by the two bonds linking the consecutive pairs of monomers indexed by $(i-1,i)$ and $(i,i+1)$, respectively. The equilibrium bond length $l_b\simeq 0.97\,\sigma$ separating two consecutive monomers finally results from the competition between WCA repulsion and an additional freely-extensible nonlinear elastic bonded potential\cite{Gres86, Krem90}; see Ref.~\onlinecite{Egor16-1} for the details of the full Hamiltonian employed. The persistence length $l_p$ of the chains is thus fully determined by $\epsilon_b$, while their contour length $l_c$ is simply related to their number $N$ of constituent monomers through $l_c \simeq (N-1) l_b$.
\par
In the framework of Sec.~\ref{subsec:DFT_FY}, the Mayer $f$-function relative to the pair interaction energy of two arbitrary molecules described by the respective conformations $(P_1,P_2)\in\Omega^2$ then reads as
\begin{equation*}
  f(\mathbf{r}_{12}, \mathcal{R}_1, \mathcal{R}_2) = \exp\Bigg(-\beta \sum_{i\in P_1} \sum_{j\in P_2} u^{\rm WCA}_{ij}\Bigg) - 1,
\end{equation*}
where the double sum runs over all pairs of monomers $i\in P_1$ and $j\in P_2$, and implicitly depends on the centre-of-mass separation vector $\mathbf{r}_{12}$ and orientations $\mathcal{R}_1$, $\mathcal{R}_2$ of the two chains. Following the procedure of Sec.~\ref{subsec:DFT_procedure}, this quantity may then readily be integrated over the two-particle configurational space and ensemble-averaged over the conformational space $\Omega$, where $\Omega$ may be preliminarily determined by molecular simulations using a given set of parameters for the above potentials. In the following, we choose the model unit of energy such that $\beta \epsilon = 1$, and parameterise the Hamiltonian in terms of the chain persistence length $l_p \simeq \beta \epsilon_b l_b$\cite{Honn90}, fixing all other constants to the values reported in Ref.~\onlinecite{Egor16-1}. We further set the molecular volume $v_0$ to that of a linear chain of fused hard spheres with radius $\sigma$ and separation distance $l_b$\cite{Varg00}. Note that the latter assumption is somewhat arbitrary; however, the choice of $v_0$ is only relevant to the PL prefactor (Eq.~\eqref{eq:pl}), whose effects on the I/N transition of the systems considered here are fairly limited, and is of little consequence in our case.
\par
We here choose to evaluate the particle conformational space through basic single-chain MC simulations, using random translational moves for individual beads combined with a simple Metropolis acceptance criterion. The maximum step size was adjusted over an equilibration run of $1\times 10^8$ steps to yield an average acceptance rate of about 30\% for a given set of $l_c$ and $l_p$, starting from an initial linear conformation. Production runs of $2.5\times 10^9$ steps were performed, and molecular trajectories were constructed by appending one chain snapshot every $1\times 10^6$ iterations. The resulting set of 2500 uncorrelated conformations was then used to evaluate the relevant ensemble-averaged integrals in the UCA for each chosen chain stiffness and contour length. This ensemble is henceforth referred to by $\Omega_0$, where the subscript denotes the fact that the UCA represents the particle conformational space in the limit of vanishing polymer concentration.
\par
\begin{figure}[]
  \includegraphics[width=\columnwidth]{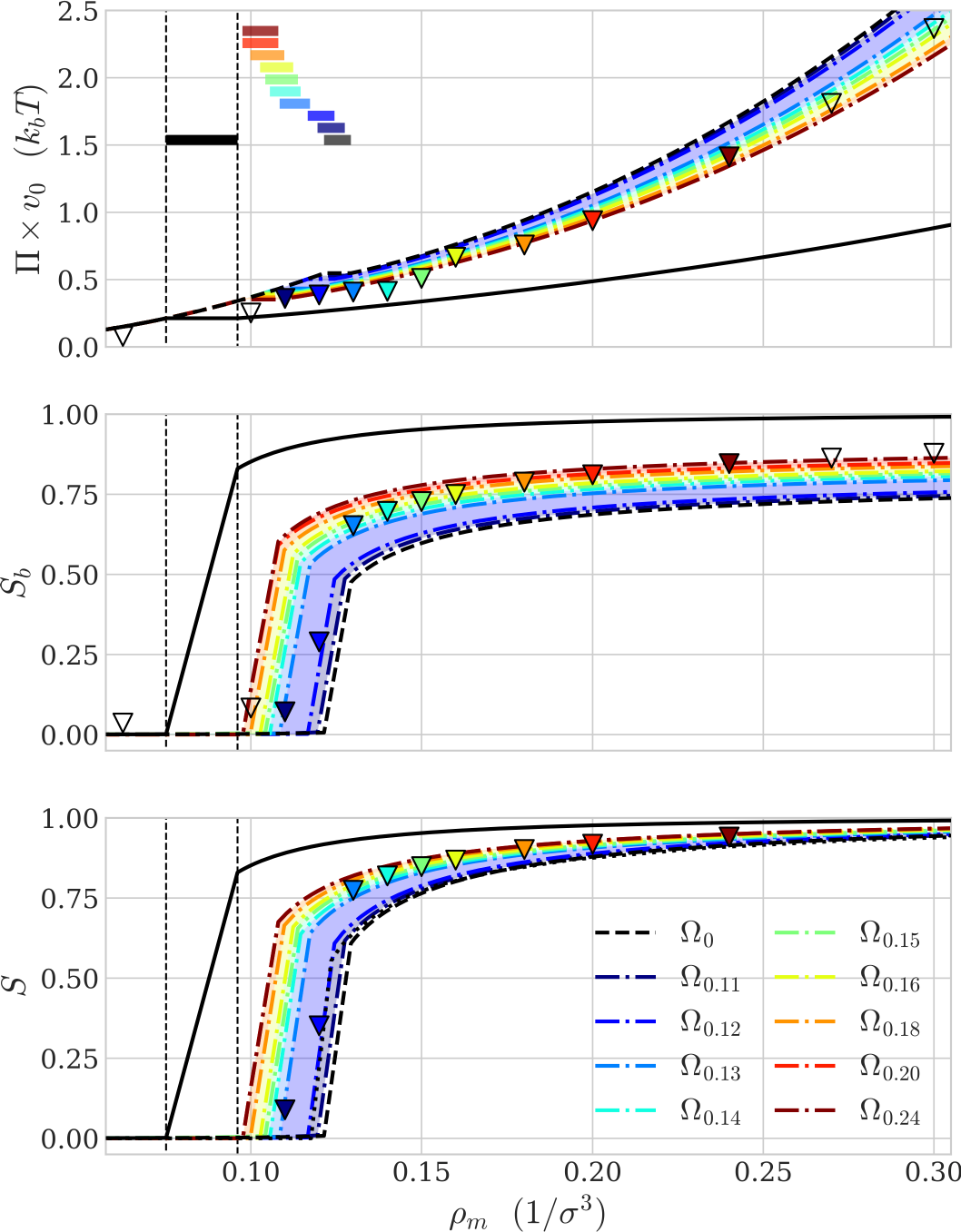}
  \caption{\label{fig2}Osmotic pressure $\Pi$, bond order parameter $S_b$ and molecular order parameter $S$ as a function of monomer number density $\rho_m$ for KG chains with persistence length $l_p = 128 \,l_b$ and $N=64$ beads (contour length $l_c = 63\,l_b$). Markers denote the simulation results of Ref.~\onlinecite{Egor16-1}, from which $S$ and $S_b$ were determined through standard means as the largest eigenvalues of the corresponding $Q$-tensors\cite{Wils96}. Black solid lines represent our DFT predictions for fully-rigid chains ($\Omega_\infty$). Black dashed lines correspond to our DFT results in the UCA ($\Omega_0$), and the black dotted line denotes the DFT data of Ref.~\onlinecite{Egor16-1} (only for $S$). Dash-dotted lines were computed using the conformational space $\Omega_{\rho}$ obtained from the bulk simulations of Ref.~\onlinecite{Egor16-1} at various fixed densities $\rho$, as described in the text. The coloured stripes in the upper panel delimit the respective I/N coexistence ranges predicted by the FY theory using the different conformational spaces, in which order parameters were computed through a standard lever-rule interpolation between the isotropic and nematic binodal points.}
\end{figure}
\begin{figure}[]
  \includegraphics[width=\columnwidth]{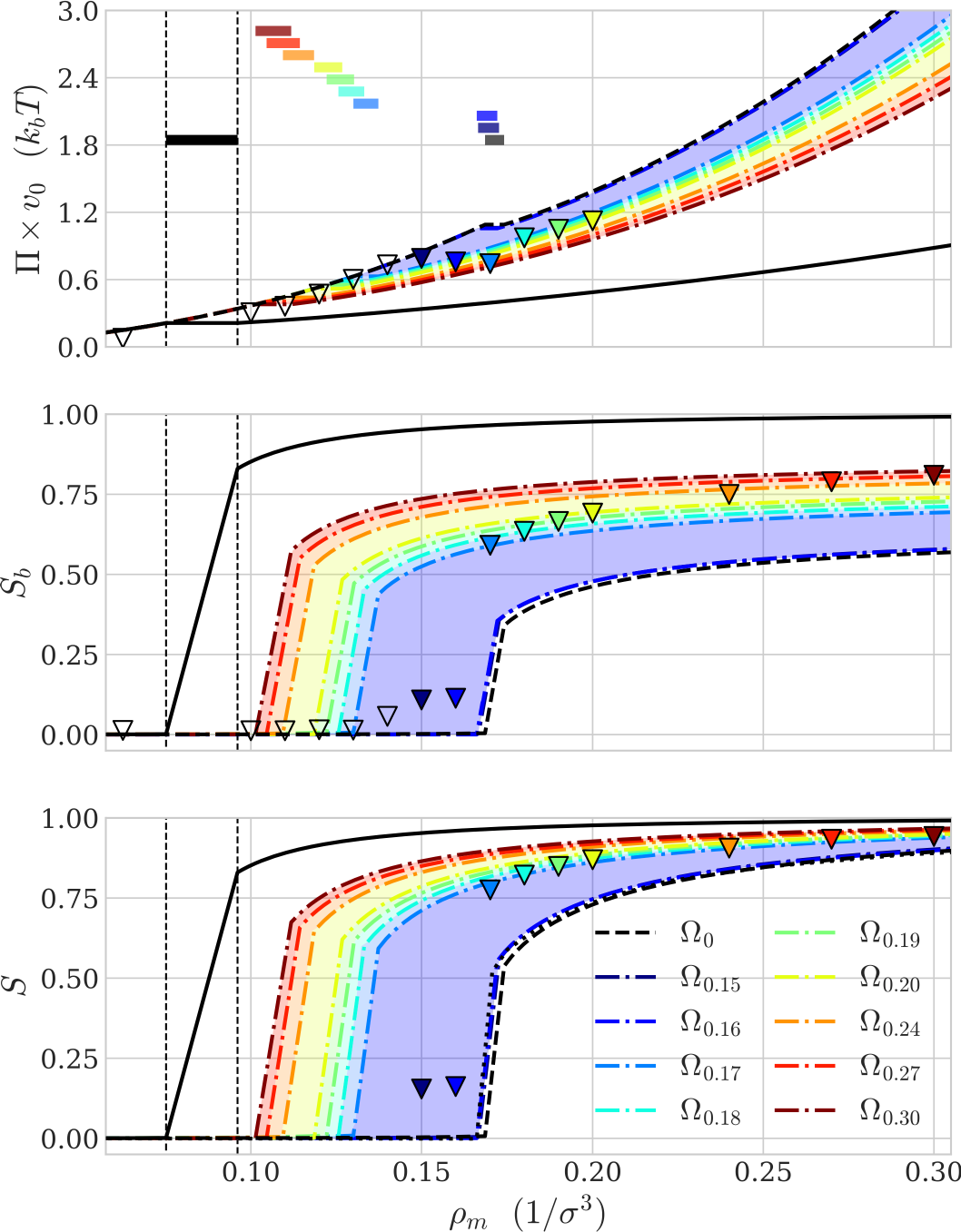}
  \caption{\label{fig3}Same as in Fig.~\ref{fig2} for chains with $l_p = 64 \,l_b$.}
\end{figure}

We summarise in Figs.~\ref{fig2} and \ref{fig3} the comparison of our DFT results with the simulation and theory data of Ref.~\onlinecite{Egor16-1} for chains comprised of $N=64$ monomers with $l_p/l_c \simeq 2$ and $1$, respectively. We first remark rather generally that the predictions of our approach in the UCA for the I/N transition density appear to be in good agreement with simulations results in both cases, with the increased particle flexibility of the systems in Fig.~\ref{fig3} deferring the I/N transition to significantly higher volume fractions. However, the further quantitative comparison of nematic properties as described by theory and simulation requires one to carefully distinguish between the different measures of orientational order underlying the FY and KS approaches. In the notation of Sec.~\ref{sec:DFT}, the dependence of the former's ODF $\psi^{\rm FY}_{\rm eq}$ on the particle long axes $\mathbf{u}$ allows for the quantification of nematic alignment about the director $\mathbf{n}$ through the usual uniaxial order parameter $S$\cite{deGe93},
\begin{equation}
  \label{eq:s_mol}
  S = 2\pi \oint d\mathbf{u} \times \frac{3(\mathbf{u}\cdot\mathbf{n})^2 - 1}{2} \times \psi^{\rm FY}_{\rm eq}(\mathbf{u}\cdot\mathbf{n}),
\end{equation}
while the contour-dependant ODF $\psi^{\rm KS}_{\rm eq}$ of the latter naturally leads to the definition of a generalised uniaxial order parameter $S_b$,
\begin{equation}
  \label{eq:s_bnd}
  S_b = 2\pi \oint d\mathbf{t} \times \frac{3(\mathbf{t}\cdot\mathbf{n})^2 - 1}{2} \times \big\langle\psi^{\rm KS}_{\rm eq}\big\rangle_c(\mathbf{t}\cdot\mathbf{n}),
\end{equation}
with $\big\langle\psi^{\rm KS}_{\rm eq}\big\rangle_c$ the contour-averaged equilibrium ODF,
\begin{equation*}
  \big\langle\psi^{\rm KS}_{\rm eq}\big\rangle_c(\mathbf{t}\cdot\mathbf{n}) = \frac{1}{l_c} \int_0^{l_c} ds \times \psi^{\rm KS}_{\rm eq}(s, \mathbf{t}\cdot\mathbf{n}).
\end{equation*}

The two order parameters $S$ and $S_b$ therefore generally correspond to rather different physical quantities, as the contour-averaged ODF $\big\langle\psi^{\rm KS}_{\rm eq}\big\rangle_c$ appearing in Eq.~\eqref{eq:s_bnd} effectively incorporates the intra-molecular dispersion of the chain segment orientations $\mathbf{t}$, which is not directly accounted for by the coarser-grained $\psi^{\rm FY}_{\rm eq}$ of Eq.~\eqref{eq:s_mol}. It is proven in Appendix~\ref{app:ord_param} that under the assumptions of the FY theory, the general relation between $S$ and $S_b$ may be derived in the compact form
\begin{equation}
  \label{eq:s_dec}
  S_b = S \times S_\Omega,
\end{equation}
with $S_\Omega$ an ensemble-averaged intra-molecular order parameter describing the distribution of chain segments $\mathbf{t}$ about the long axes $\mathbf{u}$ of the polymers as characterised by an arbitrary conformational space $\Omega$,
\begin{equation}
  \label{eq:s_int}
  S_\Omega = \bigg \langle\bigg \langle \frac{3 (\mathbf{t}\cdot \mathbf{u})^2 -1}{2}\bigg \rangle_c \bigg \rangle_\Omega.
\end{equation}
In the following, we assimilate the generic chain segment $\mathbf{t}(s)$ to the normalised bond vector $\mathbf{t}_i$ joining the monomers $i$ and $i+1$ of a discrete KG chain, and thus refer to $S$ and $S_b$ as the \textit{molecular} and \textit{bond order parameters}, respectively.
\par 
The comparison of simulation results with DFT in the UCA in terms of $S$ reveals that DFT tends to gradually underestimate the level of alignment in the nematic phase with increasing particle flexibilities, contrary to the findings of Ref.~\onlinecite{Egor16-1}. This discrepancy stems from the fact that the authors of that study actually compared the bond order parameter $S_b$ calculated from simulations with the molecular order parameter $S$ of the FY description. Our predicted values for $S$ are however found to be virtually identical to the DFT data reported in Ref.~\onlinecite{Egor16-1}, in which a simplified implementation of the FY theory was used based on an empirical expression for $\big\langle E \big\rangle_{\Omega_0}$ reported in Ref.~\onlinecite{Fyne98} for semi-flexible tangent hard-sphere chains. This observation demonstrates that the latter system provides a reliable approximation of the KG chains in this context.
\par
The DFT underestimation of nematic order can be easily understood as a direct consequence of the UCA, which does not account for the potential stretching of the chains due to the surrounding nematic field, as the particles gradually forego conformational entropy to optimise their liquid-crystalline packing\cite{Odij86}. It was thus reported in Ref.~\onlinecite{Egor16-1} that while the root-mean-square end-to-end distance $L(\rho)$ generally remains roughly constant in the isotropic phase, the I/N transition is usually associated with a rather steep jump in $L$ followed by a slower increase towards the saturation value $l_c$, as the chains progressively straighten along the nematic director. Therefore, the UCA remains quantitatively accurate up to the I/N transition, but increasingly overestimates the effects of particle flexibility as one moves deeper into the nematic range --- especially for chains with lower relative stiffnesses $l_p/l_c$, whose accessible conformations are most affected by these phase-induced restrictions\cite{Egor16-1}.
\par
The effects of this shortcoming of the FY theory may be probed by replacing the previous single-chain conformational space $\Omega_0$ obtained in the limit of infinite dilution with an ensemble $\Omega_\rho$ of single-chain conformations obtained from bulk molecular simulations at fixed finite density $\rho$. We here use for $\Omega_\rho \equiv \Omega_{\rho_m}$ a set of 30\,000 particle conformations obtained from the analysis of the simulation data of Ref.~\onlinecite{Egor16-1} at various monomer densities $\rho_m$, as indicated in Figs.~\ref{fig2} and \ref{fig3}. As expected from the previous discussion, we thus find the DFT results for $\Omega_\rho$ taken at simulation state points in the isotropic phase to be in very good agreement with those obtained using $\Omega_0$ in the UCA, which indicates the absence of significant chain stretching prior to the I/N transition for the systems considered here. 
\par
While this observation confirms the suitability of the UCA to describe the free energy of the isotropic phase in our case, the use of $\Omega_0$ in the nematic phase leads to an overestimation of the particle conformational entropy associated with an underestimation of their collective packing entropy, resulting in the observed underestimation of local nematic order. The partial compensation of these two effects in the total nematic free energy $\mathscr{F}$ may thus explain the ability of the UCA to correctly capture the rough location of the I/N transition zone, although its detailed predictions in terms of I/N coexistence densities should generally be interpreted with caution, as discussed in the next paragraph. Conversely, the use of simulated ensembles $\Omega_\rho$ obtained at nematic state points leads to increasing overestimations of chain stretching in the lower-density regions, and thus gradually pushes the I/N transition predicted by DFT to unrealistically-low concentrations for the more flexible particles.
\par
Hence, it is apparent from Figs.~\ref{fig2} and \ref{fig3} that while the use of a given ensemble $\Omega_\rho$ generally enables DFT to correctly reproduce simulation results at that same density $\rho$, owing to the state-function character of the underlying free energy $\mathscr{F}$, no single representation of the particle conformational space may yield an accurate description of thermodynamic properties at all concentrations in the framework of the FY description. Such limitations are particularly stringent for the accurate computation of the I/N coexistence range due to the abrupt density variations of the conformational space in this region, illustrated for instance by the wide gap between the DFT predictions relative to the pre-transition $\Omega_{0.16}$ and the post-transition $\Omega_{0.17}$ in Fig.~\ref{fig3}. These effects need to be taken into account for the reliable resolution of the coupled phase-coexistence equations, and may explain the extremely-narrow width of the coexistence domains predicted by DFT in Figs.~\ref{fig2} and \ref{fig3}, delimited by the plateaus in the osmotic pressure $\Pi$. However, the lack of detailed I/N coexistence analysis in the simulations of Ref.~\onlinecite{Egor16-1} precludes the thorough quantitative investigation of these considerations, whose effects on the comparison of DFT results with experimental observations are further discussed in Sec.~\ref{sec:dna}.
\par
Finally, one may investigate the impact of the coarse-grained treatment of orientational order through the calculation of the DFT-predicted bond order parameters $S_b$ from Eqs.~\eqref{eq:s_dec} and \eqref{eq:s_int}. Comparison with simulations suggests that the underestimations of $S_b$ using the UCA are more pronounced than for $S$, which could be partly attributed to the unsuitability of the FY molecular ODF to describe the local ordering of neighbouring chain segments in solution. However, the good quantitative agreement for $S_b$ achieved at given density $\rho$ through the use of $\Omega_\rho$ suggests that the main limiting factor may actually lie in the UCA itself. Indeed, it can be seen from Eq.~\eqref{eq:s_dec} that
\begin{equation*}
  S_b \leq S_\Omega,
\end{equation*}
with the equality being attained in the limit of perfect crystalline order for the particle long axes. It is shown in Appendix~\ref{app:bend_sc} that in the case of stiff unconfined chains ($l_p \gg l_c$), the intra-molecular order parameter $S_{\Omega_0}$ may be approximated as
\begin{equation}
  S_{\Omega_0} \simeq 1 - \frac{5l_c}{6l_p}.
\end{equation}
\par
The slow saturation of $S_{\Omega_0}$ with increasing persistence lengths, which reflects the presence of small undulations of the free chains about their long axes, therefore imposes an unrealistic upper bound for $S_b$ in the UCA even in the case where $l_p$ is significantly larger than $l_c$. One thus recovers the important concept of \textit{deflection length}\cite{Odij86}, denoted by $l_d$, which quantifies the average length-scale over which polymers are diverted from their unconfined conformations by the presence of the surrounding nematic field. While our results suggest that the FY molecular order parameter $S$ may be reasonably captured by the UCA for chains such that $l_p \gtrsim 2 l_c$, the accurate description of the KS bond order parameter $S_b$ in the UCA thus requires the much more stringent inequalities
\begin{equation*}
 l_c < l_d \ll l_p,
\end{equation*}
which corresponds to the so-called \textit{rigid-rod limit} where the effects of particle flexibility may be neglected altogether\cite{vanD94}. This limitation therefore requires the description of the detailed conformational statistics of polymers in a confining nematic field as a necessary prerequisite for any form of DFT based on a finer-grained representation of orientational order. This task may be further complicated by the reported importance of collective deflection fluctuation modes\cite{Egor16-1,Egor16-2}, which may not be easily captured by the mean-field treatment of KS-based descriptions, and may entail significant further theoretical developments for the reliable implementation of such approaches.

\section{The cholesteric behaviour of DNA duplexes} \label{sec:dna}
As an example application of the techniques introduced here to a more complex and experimentally-realistic system, we now turn our focus to solutions of DNA duplexes. DNA constitutes a well-studied semi-flexible biopolymer with relatively-high stiffness, whose persistence length has been measured to be 130 base pairs (\SI{44}{\nano\meter}) at salt concentration $c_\chem{Na^+} = \SI{0.5}{\molar}$\cite{Herr13}, although the precise value is somewhat dependent on solution conditions\cite{Wenn02,Herr13} and sequence\cite{Gegg10}. DNA is also chiral, both in terms of its excluded volume --- it is a grooved double helix with the major groove larger than the minor groove --- and its electrostatics --- it has a double helical pattern of negative charges associated with the phosphate groups along the backbone. Consequently, DNA duplexes in concentrated solutions generally assemble into cholesteric phases, which exhibit a rich structural polymorphism\cite{vanW90,Livo96,Naka07,Zanc10}. Highly-oriented dense DNA packings are also relevant to biology, for example, occurring in bacterial nucleoids\cite{Gour78}, dinoflagellate chromosomes\cite{Livo88} and sperm cells\cite{Blan01}.
\par
Here we will focus on DNA duplexes that are of the order of the persistence length. In particular, we will compare to the results of Refs.~\onlinecite{Strz91,Stan05}, which consider DNA duplexes with reported most-probable lengths of about 146 base pairs. As the diameter of DNA duplexes due to excluded volume is about \SI{25}{\angstrom}, these examples have an aspect ratio of about 20 --- although their effective aspect ratio in solution is expected to be somewhat lower due to the effects of electrostatics\cite{Onsa49,Stro86}. The former study focuses on the phase diagram, in particular the coexisting concentrations of the cholesteric and isotropic phases\cite{Strz91}, whereas the latter's emphasis is on the dependence of the cholesteric pitch on DNA concentration and solution conditions\cite{Stan05}. One of the particularly noteworthy results of the latter is that the cholesteric phase is left-handed with a pitch that depends non-monotonically on the DNA concentration. 
\par
The left-handed character is intriguing, as basic considerations based on the packing of hard threaded objects suggest that right-handed particles with a thread angle less than \SI{45}{\degree} (it is about \SI{30}{\degree} for DNA) should generally exhibit an entropically-stabilised right-handed phase\cite{Stra76}. This simple argument has been found to be reasonably well-obeyed for model systems with purely steric interactions\cite{Frez14,Tort17-2}. However, the addition of soft extra-molecular interactions may potentially add an additional level of complexity. If one considers the negative charges of the phosphates only, it has been postulated that the electrostatic repulsion would instead be minimised in a left-handed configuration\cite{Tomb05,Cher08}, although recent simulation results for coarse-grained particles with helical charge distributions appear to challenge this hypothesis\cite{Ruzi16,Kuhn16,Wu18}. Conversely, this preference has also been suggested to be reversed if one further considers the possibility of counterion condensation within the grooves of the helix\cite{Cher08}.
\par
To try to get a more rigorous insight into these questions, a number of theoretical studies have attempted to explain the observed experimental behaviour. Firstly, Kornyshev \textit{et al.}~exclusively considered the potential chirality arising from electrostatic interactions, and although they predicted a dependence of the pitch on concentration that resembled experiment\cite{Korn02}, it was later revealed that the corresponding pitches had in fact the wrong handedness\cite{Cher08}. Secondly, Tombolato and Ferrarini applied a DFT approach to a simplified representation of the DNA duplexes, and found that when only the steric interactions were considered, a right-handed phase resulted --- in agreement with the previous geometric argument. By contrast, when electrostatic interactions were included, this tendency was reversed and a left-handed phase was predicted, whose pitch was further reported to be in reasonable agreement with experiment for the single DNA concentration considered\cite{Tomb05}. 
\par
However, numerous approximations were introduced in this study for the computation of the virial-type integrals in Eq.~\eqref{eq:Frank}, which underpin the determination of the equilibrium pitch $\mathcal{P}_{\rm eq}$ from Eq.~\eqref{eq:pitch} in the framework of Straley's DFT\cite{Stra76}. As phase chirality is generally very weak in experimental cholesterics, DFT predictions for the pitch of such systems can be very sensitive to the accuracy of the calculations of such integrals. Thus, we have previously shown\cite{Tort17-1} that such results should generally be interpreted with caution, unless a well-defined numerical approach is used that allows for the demonstrable convergence of the procedure to the required accuracy. 
\par
Both of the above studies also treated DNA as an infinitely rigid molecule. However, flexibility is likely to have a large impact on the phase behaviour of polymers whose contour lengths are of the order of the persistence length, as is evident from Sec.~\ref{sec:nematic}. The potential effects of flexibility on the cholesteric pitch are largely unexplored\cite{Wu18}, but should be expected to be equally significant in light of the previous discussions.
\par
Here, we apply the current DFT approach to 146-base pair DNA duplexes. In order to both sample the particle conformational space and to calculate the inter-duplex interactions in the procedure of Sec.~\ref{subsec:DFT_procedure}, we make use of the oxDNA model\cite{Ould11,Snod15}. OxDNA is a nucleotide-level coarse-grained model that has been widely applied to the study of DNA-based systems, both in the context of biophysics and nanotechnology\cite{Doye13}. It provides a correct description of the mechanical properties of double-stranded DNA\cite{Roma13,Mate15}, exhibits sensible values for the persistence length and the twist modulus\cite{Snod15}, and has been shown to also capture more subtle phenomena such as twist-bend coupling\cite{Skor17}. We note that the electrostatic interactions in the model are of a simple Debye-H\"{u}ckel-like form and have been parameterised to reproduce the salt-concentration dependence of the thermodynamics of hybridisation\cite{Snod15}, rather than any specific properties related to inter-duplex interactions.
\par
We now generate the UCA ensemble $\Omega_0$ by means of single-duplex MD simulations, using an Anderson-like thermostat at temperature $T=\SI{20}{\celsius}$. Equilibration was performed over $1\times 10^7$ simulation steps, and production runs of $1 \times 10^9$ steps were used to generate a set of 2000 uncorrelated configurations. The molecular volume $v_0$ relative to the PL approximation (Eq.~\eqref{eq:pl}) was computed using the average B-DNA nucleotide volumes measured in Ref.~\onlinecite{Nada01}, and mass concentrations were obtained assuming a molar weight of \SI{650}{\dalton} per base pair.
\par
The results for the isotropic-to-cholesteric (I/C) transition densities are presented in Table~\ref{tab:coex}. As expected from Sec.~\ref{sec:nematic}, the onset of cholesteric order occurs at a much lower concentration when DNA is considered to be rigid. Significantly, the predictions when flexibility is taken into account in the UCA are in much better agreement with the experimental densities $c_c$ reported in Refs.~\onlinecite{Strz91,Stan05}, although the latter's measurements may not very accurately reflect the true cholesteric binodal point\cite{Stan05}. The dependence of the results on salt concentration is also found to follow the experimental trends, with stronger electrostatic screening generally deferring the I/C transition range to higher densities, as predicted by theory\cite{Stro86}.

\begin{table}
\caption{Isotropic/cholesteric coexistence concentrations for near-persistence-length B-DNA duplexes at $T=\SI{20}{\degree}$ and various salt concentrations. $\Omega_0$ and $\Omega_\infty$ denote the DFT predictions of this work obtained in the UCA and in the limit of infinite duplex stiffness, respectively, and are compared with the experimental results of Refs.~\onlinecite{Strz91,Stan05}. All values are reported in \SI{}{\milli\gram/\milli\litre}.}
{\begin{tabular}{l cc cc cc cc} \toprule
 & \multicolumn{2}{c}{Ref.~\onlinecite{Strz91}} & \multicolumn{2}{c}{Ref.~\onlinecite{Stan05}} & \multicolumn{2}{c}{$\Omega_\infty$} & \multicolumn{2}{c}{$\Omega_0$}  \\ \cmidrule(l{2pt}r{2pt}){2-3} \cmidrule(l{2pt}r{2pt}){4-5}  \cmidrule(l{2pt}r{2pt}){6-7}  \cmidrule(l{2pt}r{2pt}){8-9}
 $c_{\chem{Na^+}}$& $c_i$ & $c_c$ & $c_i$ & $c_c$ & $c_i$ & $c_c$ & $c_i$ & $c_c$ \\ \midrule
 \SI{0.5}{\molar} & n/a & n/a & n/a & 218 & 120 & 138 & 242 & 250 \\
 \SI{1.0}{\molar} & 171 & 270 & n/a & 257 & 132 & 155 & 268 & 276 \\ \bottomrule
\end{tabular}}
\label{tab:coex}
\end{table}

One notable difference with the results of Ref.~\onlinecite{Strz91} lies in the width of the coexistence region, which is found to be strongly underestimated by DFT. This discrepancy may be partially attributable to the shortcomings of the UCA in the vicinity of the I/C transition, as discussed in Sec.~\ref{sec:nematic}, and could also reflect potential inaccuracies of the oxDNA model. Furthermore, our theory does not account for the substantial polydispersity reported in the experimental systems\cite{Strz91}, which should be expected to significantly widen the phase coexistence range\cite{Vroe92,Sato96}.
\par
\begin{figure}[]
  \includegraphics[width=\columnwidth]{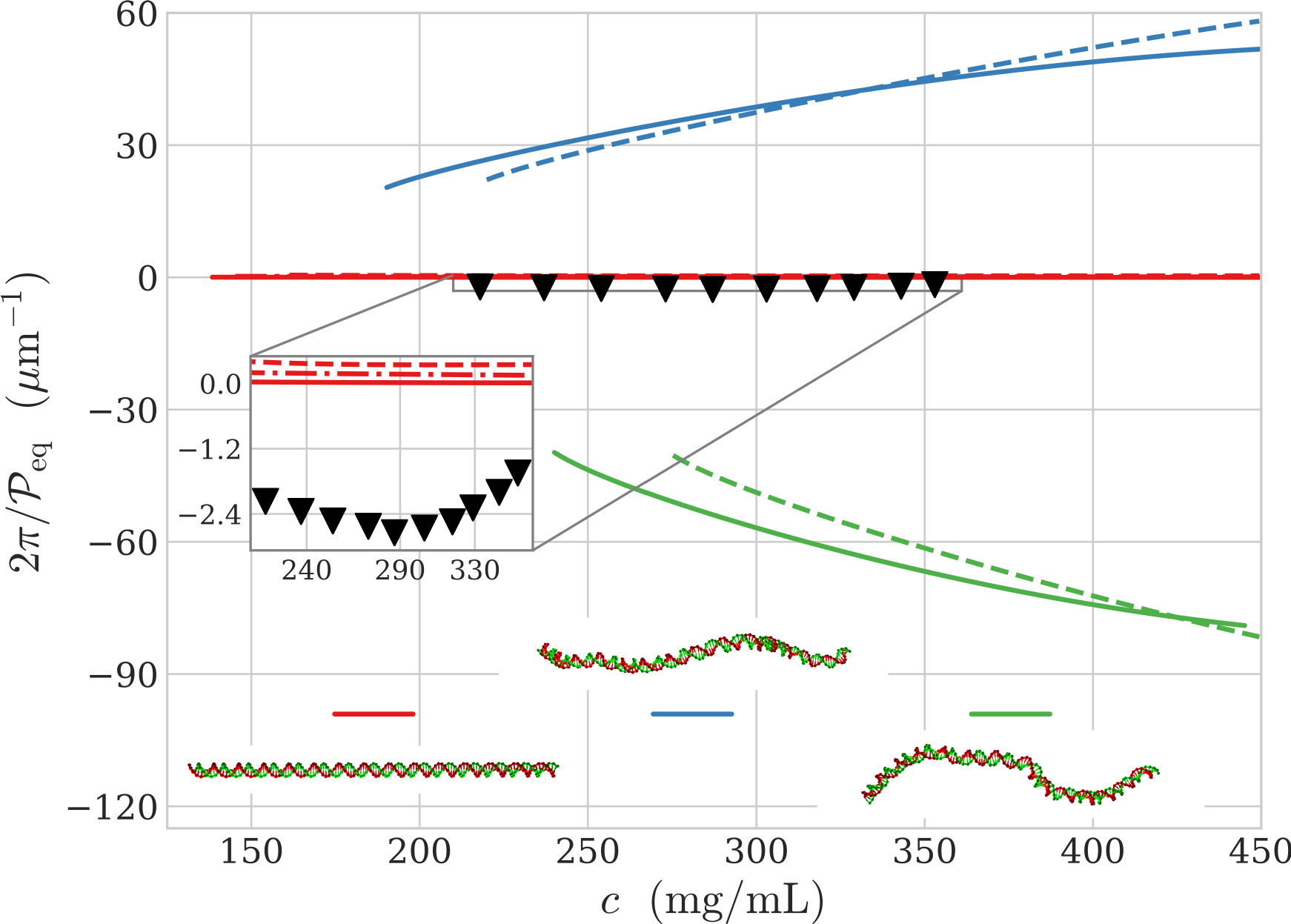}
  \caption{\label{fig4}Inverse equilibrium cholesteric pitch $2\pi/\mathcal{P}_{\rm eq}$ as predicted by DFT for 146 base-pair DNA duplexes in the limit of infinite particle stiffness ($\Omega_\infty$, in red) and for two instantaneous conformations, respectively observed to possess a strong left- (blue) and right-handed (green) character. Solid lines were computed using the oxDNA-parameterised excluded volume and Debye-H\"uckel interactions at $T=\SI{20}{\degree}$ and $c_{\chem{Na^+}} \! = \SI{0.5}{\molar}$. The dash-dotted line was similarly calculated by replacing the Debye-H\"uckel contribution with the electrostatic potential of Ref.~\onlinecite{Tomb05} (only for $\Omega_\infty$). Dashed lines were obtained in the limit of pure steric repulsion ($c_{\chem{Na^+}} \!\to \infty$). Markers denote the experimental results of Ref.~\onlinecite{Stan05}. The lowest densities reported correspond to the cholesteric binodal points $c_c$ predicted by DFT.}
\end{figure}

We now turn our focus to the cholesteric pitch, and first consider rigid straight duplexes in order to compare our results with those previous theoretical investigations. As illustrated in Fig.~\ref{fig4}, we find the predicted pitch to be right-handed when we do not include the electrostatic term in the potential, as expected from the previous excluded volume argument. Interestingly, the magnitude of the pitch is also found to be much larger than the experimental values, indicating a significant underestimation of phase chirality. However, unlike in Ref.~\onlinecite{Tomb05}, we do not report evidence of a handedness inversion when electrostatic repulsion is taken into account, but rather observe a further reduction of the predicted phase chirality. 

This observation could potentially be explained by the competition between the entropic favourability of right-handed duplex configurations due to excluded-volume contributions and the hypothetical enthalpic preference for left-handed arrangements to minimise electrostatic repulsion, as postulated in Ref.~\onlinecite{Tomb05}. However, we find a more likely explanation to lie in the fact that electrostatics penalise the close-pair configurations where the chirality of the excluded volume is most relevant, so that longer-range repulsion simply `smears out' the local details of the chiral surface structure of DNA. To verify that this effect is not simply an artefact of our chosen description of electrostatic contributions in the oxDNA model, we repeated the calculations using the electrostatic potential detailed in Ref.~\onlinecite{Tomb05}. Our results were however found to be very similar in both cases, which leads to the conclusion that the prediction of a left-handed phase in Ref.~\onlinecite{Tomb05} is most likely due to some of the numerical approximations introduced in the virial calculations therein.
\par
We have also attempted to compute the cholesteric pitch including the effects of flexibility in the UCA. However, we concluded after extensive calculations that the drastic dependence of the predicted pitches on the sampling of the particle conformational space precluded their determination with sufficient statistical accuracy. To illustrate the nature of the problems, we report in Fig.~\ref{fig4} the equilibrium pitches computed for two example conformations taken from $\Omega_0$, both of which were observed to possess noticeably helicoidal shapes with opposite respective handednesses. Interestingly, the handedness of the equilibrium pitch is then obtained to be opposite to that of the corresponding helicoidal conformation, which mirrors theoretical predictions for hard, weakly-curled coarse-grained helices\cite{Frez14,Duss15,Tort17-1}. Furthermore, the phase chirality is found to be considerably stronger for these conformations, and the resulting pitches only weakly depend on the inclusion of electrostatics,  as the length-scales associated with the shape helicity of the duplexes is in this case considerably larger than that of their screened electrostatic interactions.
\par
The strong variations in the predicted pitches between individual DNA conformations prevented the reliable convergence of the DFT procedure, and the large fluctuations in both their handedness and magnitude obtained using different independent ensembles $\Omega_0$ did not allow for the observation of any meaningful tendencies. Therefore, we could not reliably determine whether such helical conformations may be linked to the underlying intra-molecular chirality of DNA, nor what their net effect on the cholesteric pitch should be, given a large enough statistical sample of the conformational space. Moreover, it is apparent from Sec.~\ref{sec:nematic} that the UCA should generally be unsuitable for the description of the nematic --- and \textit{a fortiori} cholesteric --- phases of particles whose contour lengths are of the order of the persistence length, so we made no attempt to pursue this study further.
\par
It nonetheless transpires that the flexibility of DNA duplexes plays a significant role in their liquid-crystalline assembly, which cannot be reasonably neglected in any theoretical investigation of their phase behaviour --- especially in the context of the fine determination of their cholesteric pitch. However, the quantitative examination of its effects is an arduous task, as near-persistence-length duplexes are too flexible to be accurately described by the FY DFT in the UCA, while the cholesteric ordering of their shorter counterparts generally entails more involved living polymerisation mechanisms~\cite{Naka07}, which further complicate their theoretical treatment~\cite{DeMi16}.
\par
In the current framework, one possible way to circumvent these issues would of course be to use the oxDNA model to simulate a nematic phase of elongated DNA duplexes at the concentration of interest, in order to generate ensembles of conformations that are more representative of a real cholesteric phase to be input into the FY DFT. However, such simulations may prove to be quite computationally challenging, although they could provide an interesting avenue for future research.

\section{Conclusion} \label{sec:conclusion}
We have investigated in detail the application of the FY DFT to the description of the nematic self-assembly of various semi-flexible particle models, and showed for the first time how it can be straightforwardly extended to account for the effects of particle flexibility on the cholesteric phase. Extensive comparisons with the results of molecular simulations of persistent KG chains reveal that the I/N transition densities of polymers with $l_p/l_c \gtrsim 1$ appear to be well captured by the theory in the UCA, although the detailed mechanisms underlying these predictions are not fully clear. This limitation stems from the fact that while the FY DFT is found to be quantitatively accurate in the isotropic phase, where no concentration-dependent stretching of the chains is observed for the systems considered here, the reliable computation of nematic properties beyond the I/N transition generally requires a more involved account of the restrictions of the accessible particle conformational space induced by the confining effects of the surrounding nematic field.
\par
These considerations deserve to be investigated further, and dedicated numerical studies of the I/N coexistence properties of persistent chains, along the lines of the simulations of Ref.~\onlinecite{Dijk95} for systems of jointed sphero-cylinders, would be desirable. Moreover, the effects of these restrictions should generally depend on the scaled contour length $l_c/\sigma$ as well as $l_p/l_c$, while the former ratio was kept fixed in this paper. However, the quantitative agreement between FY DFT predictions and simulation results reported in Ref.~\onlinecite{Egor16-1} was found to be relatively unaltered by varying of the contour length in the range $l_c/\sigma \in [15,63]$ at given $l_p/l_c$, so it is expected that the previous discussions should quite generally apply to elongated semi-flexible particles.
\par
We additionally demonstrate that the use of a molecular-level representation of orientational order allows the FY theory to circumvent some of these shortcomings by coarse-graining over conformational fluctuations on the scale of the deflection length, which may provide for a satisfactory description of the nematic phase of polymer chains with $l_p/l_c\gtrsim 2$. While this statement may be somewhat limiting in scope, we feel that such a restriction is largely inherent to the underlying formalism of the Onsager DFT, as the complex coupling between conformational statistics and nematic order is further supplemented by the growing inadequacy of the second-virial approximation to describe the higher-density I/N transitions associated with increased particle flexibilities. It is also worth remarking that previous numerical studies have demonstrated the quantitative accuracy of the FY approach to consistently match (and in many cases surpass) that of other DFT implementations across a wide range of chain contour lengths $l_c$ and relative persistence lengths $l_p/l_c$\cite{Fyne98,Jaff01,Egor16-1}, which points to the rather general unsuitability of existing theoretical frameworks to tackle polymeric systems of intermediate stiffness\cite{Vroe92}.
\par
Such limitations are particularly stringent for the determination of cholesteric properties, which are especially sensitive to the detailed nature of the conformational fluctuations undergone by flexible particles. It is thus found that while the combination of the FY DFT and the oxDNA coarse-grained model provides for a satisfactory prediction of the I/C transition of near-persistence-length DNA duplexes, the strong dependence of the predicted pitches on minute structural changes largely precludes their realistic determination in the context of the UCA. 
\par
This analysis nonetheless reveals that steric and electrostatic interactions between rigid linear DNA duplexes appear to be unable to account for either the handedness or the magnitude of the phase chirality observed in experiments. This conclusion concurs with the recent findings of Ref.~\onlinecite{Cort17}, in which extensive all-atom MD simulations of short DNA oligomers demonstrated the absence of a significant chiral contribution attributable to inter-duplex electrostatic interactions --- thus suggesting the need for alternative mechanisms to explain their cholesteric behaviour.
\par
The accurate description of the conformational statistics of DNA duplexes in dense phases therefore emerges as a crucial prerequisite to any realistic investigation of their cholesteric assembly, and likely lies beyond the reasonable reach of current theoretical and numerical tools. Perhaps surprisingly given their considerably larger sizes, elongated DNA origamis, whose cholesteric behaviour has been recently explored experimentally\cite{Siav17}, may present a more tractable target for the application of the FY DFT due to their much greater relative stiffness. Although the persistence length of the studied 6-helix-bundle origamis is not yet as well-characterised as that of single duplexes, preliminary measurements have indeed suggested that $l_p \geq 5\, l_c$\cite{Kaue11,Schi13}. Furthermore, their aspect ratio of about 70\cite{Siav17} leads to the formation of a stable cholesteric phase in a regime approaching the Onsager limit of high dilutions\cite{Stra73-2}. 
\par
We thus expect such origamis to provide a much more suitable model system in regard to the fundamental assumptions of the FY DFT. The results of their study will be presented elsewhere.

\begin{acknowledgments}
The authors are deeply grateful to A.~Milchev for kindly sharing his simulation data. This project has received funding from the European Union's Horizon 2020 research and innovation programme under the Marie Sk\l{}odowska-Curie Grant Agreement No.~641839. The authors would like to acknowledge the use of the University of Oxford Advanced Research Computing (ARC) facility in carrying out this work. http://dx.doi.org/10.5281/zenodo.22558. We are grateful to the UK Materials and Molecular Modelling Hub for computational resources, which is partially funded by EPSRC (EP/P020194/1).
\end{acknowledgments}

%----------------------------------------------------------------------------------------
%	REFERENCE LIST
%----------------------------------------------------------------------------------------

\bibliography{molphys}

%----------------------------------------------------------------------------------------
%	APPENDICES
%----------------------------------------------------------------------------------------
\appendix

\section{Molecular (FY) and bond (KS) order parameters} \label{app:ord_param}

In the case of a discretised KG polymer chain, the formal definition of the KS nematic order parameter (Eq.~\eqref{eq:s_bnd}) may be recast in the form
\begin{equation}
  S_b \equiv \bigg\langle\frac{3(\mathbf{t}_{i}\cdot \mathbf{n})^2-1}{2} \bigg\rangle_{\{\psi_{i}\}} \equiv \bigg\langle\frac{3\cos^2\theta_{l,i}-1}{2} \bigg\rangle_{\{\psi_{i}\}},
\end{equation}
where we used the notation of Fig.~\ref{fig1}, denoting by the brackets an average over all bond orientations $\mathbf{t}_{i}$ about the uniform director $\mathbf{n}$. Such orientations may be fully characterised by a set of contour-dependent ODFs $\{\psi_{i}\}$ for all bonds $i$ comprising the chain. Neglecting the influence of the surrounding nematic field on the conformations of individual particles, one may write\cite{vanW12,vanW13}
\begin{equation}
  \label{eq:proba}
\psi_{i}(\mathbf{t}_{i}) = \psi(\mathcal{R}) \times P_{i}(\mathcal{R}^{\sf T}\cdot \mathbf{t}_{i}),
\end{equation}
with $\psi$ the molecular (FY) ODF quantifying the probability of finding a chain with orientation $\mathcal{R}$ as defined in Fig.~\ref{fig1}, and $P_{i}$ the probability of finding the $i$-th bond vector pointing along $\mathbf{t}_{i}$ in the molecular frame $\mathcal{R}$ of an arbitrary chain. Under these assumptions, one readily obtains
\begin{equation}
  \big\langle \cdot \big\rangle_{\{\psi_{i}\}} = \big\langle\big\langle\cdot \big\rangle_\psi\big\rangle_{\{P_{i}\}},
\end{equation}
where the inner brackets denote a thermodynamic average over all molecular orientations $\mathcal{R}$, and the outer brackets an ensemble average over all accessible bond orientations $\mathbf{t}_{i}$ as expressed in the molecular frame. In the FY decoupling approximation (Eq.~\eqref{eq:proba}), these two averages may thus be computed independently by projecting the normalised bond vectors $\mathbf{t}_{i}$ onto $\mathcal{R}$,
\begin{align}
\mathcal{R}^{\sf T} \cdot \mathbf{t}_{i} &= \mathcal{R}^{\sf T} \cdot  
\begin{bmatrix}
\sin\theta_{l,i}\cos\phi_{l,i} \\
\sin\theta_{l,i}\sin\phi_{l,i} \\
\cos\theta_{l,i}
\end{bmatrix}_{\mathcal{R}_{\rm lab}}  
\nonumber \\ &\equiv
\begin{bmatrix}
\sin\theta_{m,i}\cos\phi_{m,i} \\
\sin\theta_{m,i}\sin\phi_{m,i} \\
\cos\theta_{m,i}
\end{bmatrix}_{\mathcal{R}} 
,
\end{align}
with $\phi_{l,i}$ and $\phi_{m,i}$ the azimuthal bond angles in the laboratory and molecular frames, respectively. Hence,
\begin{align}
\label{eq:rotation}
\begin{bmatrix}
\sin\theta_{l,i}\cos\phi_{l,i} \\
\sin\theta_{l,i}\sin\phi_{l,i} \\
\cos\theta_{l,i}
\end{bmatrix}_{\mathcal{R}_{\rm lab}}
&=
\mathcal{R} \cdot 
\begin{bmatrix}
\sin\theta_{m,i}\cos\phi_{m,i} \\
\sin\theta_{m,i}\sin\phi_{m,i} \\
\cos\theta_{m,i}
\end{bmatrix}_{\mathcal{R}},
\end{align}
with $\mathcal{R}$ being represented in $\mathcal{R}_{\rm lab}$ by the Euler angle triad $(\alpha,\theta,\phi)$ in the $z$-$y$-$z$ convention,
\begin{widetext}
\begin{align}
\label{eq:frame}
\mathcal{R} &= 
\begin{bmatrix}
\cos\alpha\cos\theta\cos\phi - \sin\alpha\sin\phi & -\sin\alpha\cos\theta\cos\phi - \cos\alpha\sin\phi & \sin\theta\cos\phi \\
\cos\alpha\cos\theta\sin\phi + \sin\alpha\cos\phi & -\sin\alpha\cos\theta\sin\phi + \cos\alpha\cos\phi & \sin\theta\sin\phi \\
-\cos\alpha\sin\theta & \sin\alpha\sin\theta & \cos\theta
\end{bmatrix}_{\mathcal{R}_{\rm lab}}.
\end{align}
\end{widetext}
Plugging Eq.~\eqref{eq:frame} into Eq.~\eqref{eq:rotation} and projecting the result onto $\mathbf{e}_z$ leads to
\begin{multline}
  \cos\theta_{l,i} = \cos\theta_{m,i}\cos\theta + \sin\theta_{m,i}\sin\theta \\ \times (\sin\phi_{m,i}\sin\alpha-\cos\phi_{m,i}\cos\alpha),
\end{multline}
which yields, after some rearrangements,
\begin{align}
  \label{eq:c2theta}
  \cos^2\theta_{l,i} =\,& \cos^2\theta_{m,i} \cos^2\theta  \nonumber \\ +& \sin^2\theta_{m,i} \sin^2\phi_{m,i}\times\sin^2\theta\sin^2\alpha \nonumber \\+& \sin^2\theta_{m,i} \cos^2\phi_{m,i}\times\sin^2\theta\cos^2\alpha \nonumber \\ +& \frac{\sin2\theta_{m,i} \sin2\theta}{2} (\sin\phi_{m,i}\sin\alpha-\cos\phi_{m,i}\cos\alpha) \nonumber \\-& \sin2\phi_{m,i}\sin^2\theta_{m,i}\times \sin^2\theta \sin\alpha \cos\alpha.
\end{align}
\par
Let us first perform the average of Eq.~\eqref{eq:c2theta} over the molecular ODF $\psi$, which respect to which the angles $\theta_{m,i}$ and $\phi_{m,i}$ are taken to be invariant following the previous discussion. One must therefore have, for any functions $f(\theta_{m,i},\phi_{m,i})$ and $g(\mathcal{R})$,
\begin{align}
  \label{eq:indep}
  \big\langle f(\theta_{m,i},\phi_{m,i}) g(\mathcal{R}) \big\rangle_\psi &=  f(\theta_{m,i},\phi_{m,i}) \big\langle g(\mathcal{R}) \big\rangle_\psi \nonumber \\&\equiv f(\theta_{m,i},\phi_{m,i}) \oint d\mathcal{R} \times \psi(\mathcal{R}) g(\mathcal{R}).
\end{align}
Furthermore, assuming the molecular ODF $\psi$ to be cylindrically-symmetric about $\mathbf{n}$ yields
\begin{equation}
  \psi(\mathcal{R}) = \psi(\cos\theta),
\end{equation}
so that for any function $h(\theta)$,
\begin{align}
  \label{eq:uniaxial}
  \big\langle h(\theta) \big\rangle_\psi &=  \int_0^{2\pi}d\alpha \int_0^{2\pi} d\phi \int_0^\pi d\theta \sin\theta \times \psi(\mathcal{R}) h(\theta) \nonumber \\ &= 4\pi^2 \int_0^\pi d\theta \sin\theta \times \psi(\cos\theta) h(\theta).
\end{align}

Using Eqs.~\eqref{eq:indep} and \eqref{eq:uniaxial}, one obtains
\begin{widetext}
\begin{align}
    \big\langle \sin^2\theta_{m,i} \sin^2\phi_{m,i}\times\sin^2\theta\sin^2\alpha \big\rangle_\psi &= \sin^2\theta_{m,i} \sin^2\phi_{m,i} \int_0^{2\pi}d\alpha \times \sin^2\alpha \int_0^{2\pi} d\phi \int_0^\pi d\theta \sin\theta \times \psi(\cos\theta)\sin^2\theta \nonumber \\
      &=\frac{\sin^2\theta_{m,i} \sin^2\phi_{m,i} \big\langle \sin^2\theta\big\rangle_\psi}{2}, \\
  \big\langle \sin^2\theta_{m,i} \cos^2\phi_{m,i}\times\sin^2\theta\cos^2\alpha \big\rangle_\psi &= \frac{\sin^2\theta_{m,i} \cos^2\phi_{m,i} \big\langle \sin^2\theta\big\rangle_\psi}{2},
\end{align}
\end{widetext}
and it is easy to show that all the terms in the last two lines of Eq.~\eqref{eq:c2theta}, involving only odd powers of $\cos\alpha$ and $\sin\alpha$, average out to zero. Thus,
\begin{equation}
  \big\langle\cos^2\theta_{l,i} \big\rangle_\psi = \cos^2\theta_{m,i}  \big\langle \cos^2\theta \big\rangle_\psi + \frac{\sin^2\theta_{m,i}  \big\langle \sin^2\theta \big\rangle_\psi}{2}.
\end{equation}
Finally, using the shorthand 
\begin{equation}
\label{eq:p_ave}
\big\langle \cdot \big\rangle_P \equiv \big \langle \cdot \big\rangle_{\{P_{i}\}} 
\end{equation}
for the conformational average, the KS order parameter $S_b$ may be written as
\begin{widetext}
\begin{align}
  S_b &= \frac{3\big\langle\big\langle\cos^2\theta_{l,i} \big\rangle_\psi\big\rangle_P-1}{2} \nonumber \\
         &= \frac{3 \big\langle \cos^2\theta_{m,i} \big\rangle_P\big\langle \cos^2\theta \big\rangle_\psi}{2} + \frac{3\big\langle \sin^2\theta_{m,i}\big\rangle_P \big\langle \sin^2\theta \big\rangle_\psi}{4} - \frac{1}{2} \nonumber \\
         &= \frac{3 \big\langle \cos^2\theta_{m,i} \big\rangle_P\big\langle \cos^2\theta \big\rangle_\psi}{2} + \frac{3 \Big(\big\langle\sin^2\theta_{m,i}\big\rangle_P-1\Big)\Big(\big\langle\sin^2\theta \big\rangle_\psi -1\Big)}{4} +\frac{3\Big(\big\langle\sin^2\theta_{m,i}\big\rangle_P+\big\langle\sin^2\theta \big\rangle_\psi \Big)}{4} - \frac{5}{4} \nonumber \\
         &= \frac{9 \big\langle \cos^2\theta_{m,i} \big\rangle_P\big\langle \cos^2\theta \big\rangle_\psi}{4} - \frac{3\Big(  \big\langle \cos^2\theta_{m,i} \big\rangle_P + \big\langle \cos^2\theta \big\rangle_\psi \Big)}{4} + \frac{1}{4} \nonumber \\
         &= \Bigg(\frac{3\big\langle \cos^2\theta_{m,i} \big\rangle_P-1}{2}\Bigg) \times \Bigg( \frac{3\big\langle \cos^2\theta \big\rangle_\psi-1}{2} \Bigg).
\end{align}
\end{widetext}
Thus,
\begin{equation}
S_b = S \times S_\Omega,
\end{equation}
with $S$ the FY nematic order parameter quantifying the angular distribution of the molecular long axes, 
\begin{equation}
    S = \frac{3\big\langle \cos^2\theta \big\rangle_\psi-1}{2},
\end{equation}
and $S_\Omega$ an ensemble-averaged order parameter describing the distribution of bond orientations about the long axis of each chain conformation,
\begin{equation}
  \label{eq:sm}
   S_\Omega = \frac{3\big\langle \cos^2\theta_{m,i} \big\rangle_P-1}{2},
\end{equation}
which yield the discretised versions of Eqs.~\eqref{eq:s_mol}--\eqref{eq:s_int}.

\section{Bending fluctuations and intra-molecular order parameter for unconfined chains} \label{app:bend_sc}
Let us now work out the scaling behaviour of the order parameter $S_{\Omega_0}$ for unconfined chains as a function of their contour and persistence length $l_c$ and $l_p$. In the KG bead-spring model, chain stiffness is governed by the bond bending potential,
\begin{equation}
  \label{eq:bend_disc}
  U_{\rm bend} = \sum_{i=1}^{N-2} \epsilon_b \big(1 - \mathbf{t}_{i+1}\cdot\mathbf{t}_{i}\big) = \frac{\epsilon_b}{2} \sum_{i=1}^{N-2} \big( \mathbf{t}_{i+1} - \mathbf{t}_{i}\big)^2,
\end{equation}
with $\mathbf{t}_{i}$ the normalised bond vector linking the monomers $i$ and $i+1$, respectively located in $\mathbf{r}_i$ and $\mathbf{r}_{i+1}$,
\begin{equation}
  \label{eq:u_disc}
  \mathbf{t}_{i} \equiv \frac{\mathbf{r}_{i+1} - \mathbf{r}_i}{\lVert \mathbf{r}_{i+1} - \mathbf{r}_i \rVert},
\end{equation}
where $\lVert \cdot \rVert$ is the Euclidean norm. Let us denote by $l_b$ the bond length separating any two adjacent monomers, taken to be a constant, and define $s\equiv l_b \times i$ as the curvilinear abscissa of the $i$-th monomer. In the continuum limit ($l_c \gg l_b$), one may assume $\big\{\mathbf{r}(s),\mathbf{t}(s)\big\} \equiv \big\{\mathbf{r}_i,\mathbf{t}_{i} \big\}$ to be differentiable functions of the continuous variable $s$, so that Eq.~\eqref{eq:bend_disc} converges towards the Riemann integral
\begin{equation}
  \label{eq:u_cont}
  U_{\rm bend} = \frac{\epsilon_b l_b}{2} \int_0^{l_c} ds \times \bigg ( \frac{d\mathbf{t}}{ds} \bigg)^2,
\end{equation}
and Eq.~\eqref{eq:u_disc} takes the form of the simple differential
\begin{equation}
  \label{eq:diff}
  \mathbf{t} = \frac{d\mathbf{r}}{ds},
\end{equation}
with $\mathbf{r}(s)$ the continuous curve describing the chain conformation in space. One thus recovers the standard worm-like chain model, with bending stiffness $K = \epsilon_b l_b \equiv l_p k_bT$.
\par
Let us denote by $\mathcal{R} \equiv \begin{bmatrix}\mathbf{v}\times\mathbf{u} & \mathbf{v} & \mathbf{u}\end{bmatrix}$ the three unit vectors defining the molecular frame $\mathcal{R}$. In the framework of PCA, $\mathbf{v}$ and $\mathbf{u}$ respectively correspond to the molecular short and long axes, and are defined such that the following inequalities are always verified,
\begin{equation}
  \label{eq:ineq}
    \lambda^2_{\mathbf{u}} > \lambda^2_{\mathbf{v}\times\mathbf{u}} > \lambda^2_{\mathbf{v}},
\end{equation}
where the $\lambda^2_{\mathbf{x}}$ represent the principal moments of the gyration tensor, and quantify the extent of the chain along axis $\mathbf{x}$. Their expression reads, in our notation,
\begin{equation}
  \label{eq:principal}
  \lambda^2_{\mathbf{x}} \equiv \big\langle (\mathbf{r}\cdot \mathbf{x})^2 \big\rangle_c\qquad \forall \, \mathbf{x} \in \{\mathbf{v}\times\mathbf{u},\mathbf{v},\mathbf{u}\},
\end{equation}
with $\big\langle \cdot \big\rangle_c$ the contour average,
\begin{equation}
  \label{eq:contour}
   \big\langle \cdot \big\rangle_c \equiv \frac{1}{l_c} \int_0^{l_c} ds \: \cdot \:,
\end{equation}
assuming the particle centre of mass to be set to the origin of the frame, i.e., $\big\langle  \mathbf{r}\big\rangle_c = \mathbf{0}$.
The dimensionless molecular anisotropy parameter $\kappa$ is then defined as\cite{Theo85}
\begin{equation}
  \kappa \equiv \frac{3}{2} \frac{\lambda^4_{\mathbf{u}} + \lambda^4_{\mathbf{v}} + \lambda^4_{\mathbf{v}\times\mathbf{u}}}{(\lambda^2_{\mathbf{u}} +\lambda^2_{\mathbf{v}} + \lambda^2_{\mathbf{v}\times\mathbf{u}})^2} - \frac{1}{2},
\end{equation}
which verifies $0 \leq \kappa \leq 1$, $\kappa=0$ being reached in the limit of a spherically-symmetric monomer distribution and $\kappa=1$ in the case of an ideal linear chain.
\par
In the following, we restrict our study to systems of stiff linear polymers such that $l_p \gg l_c$, for which $\kappa \sim 1$. In this case, the first inequality in Eq.~\eqref{eq:ineq} becomes wide, so that
\begin{equation}
  \lambda^2_{\mathbf{u}} \gg \lambda^2_{\mathbf{v}\times\mathbf{u}}, \lambda^2_{\mathbf{v}}.
\end{equation}
One may then approximate the end-to-end separation vector by its projection onto $\mathbf{u}$, 
\begin{equation}
  [\mathbf{r}(l_c)-\mathbf{r}(0)] \cdot \mathbf{u} = \int_0^{l_c} ds \times \mathbf{t}(s)\cdot \mathbf{u} = \int_0^{l_c} ds \times\cos \theta_m(s),
\end{equation}
where we used Eq.\eqref{eq:diff}, denoting by $\theta_m(s)$ the continuous limit of the bond angle $\theta_{m,i}$ as defined in Fig.~\ref{fig1}. The square end-to-end distance $L^2$ thus reads as
\begin{equation}
  \label{eq:l2}
  L^2 \simeq  \iint_0^{l_c} ds ds' \times \cos \theta_m(s) \cos \theta_m(s').
\end{equation}
Assuming $\theta_m$ to be a slowly-varying function of $s$ for stiff polymers, one may write
\begin{equation}
  \cos \theta_m(s') \simeq \cos \theta_m(s) + (s'-s) \frac{d\cos\theta_m(s)}{ds},
\end{equation}
so that Eq.~\eqref{eq:l2} may be recast in the form
\begin{equation}
  \label{eq:l2_2}
  L^2 \simeq l^2_c\big \langle \cos^2 \theta_m \rangle_c + \iint_0^{l_c} ds ds' \times \frac{s'-s}{2} \frac{d \cos^2  \theta_m(s)}{ds}. 
\end{equation}
The integration by parts of the last term in Eq.~\eqref{eq:l2_2} leads to, after rearrangements,
\begin{equation}
  L^2  \simeq \frac{l^2_c}{2} \bigg\{ 3\big \langle \cos^2 \theta_m \rangle_c - \frac{\cos^2\theta_m(0) + \cos^2\theta_m(l_c)}{2} \bigg\},
\end{equation}
which yields the mean square end-to-end distance $\big \langle L^2 \big\rangle_\Omega$,
\begin{equation}
 \label{eq:l2_3}
 \frac{\big \langle L^2\big \rangle_\Omega}{ l^2_c}  \simeq \frac{ 3\big \langle\big \langle \cos^2 \theta_m \big\rangle_c\big \rangle_\Omega - \big \langle\cos^2\theta_m(0)\big \rangle_\Omega}{2},
\end{equation}
where we used the equivalence of the two chain extremities. Eqs.~\eqref{eq:s_int} and~\eqref{eq:l2_3} thus provide a first expression for $S_\Omega$,
\begin{equation}
  \label{eq:s_eq}
  S_{\Omega} \simeq \frac{\big \langle L^2\big \rangle_\Omega}{l^2_c} + \frac{ \big \langle\cos^2\theta_m(0)\big \rangle_\Omega-1}{2}.
\end{equation}
\par
For inextensible and unconfined worm-like chains as described by Eqs.~\eqref{eq:u_cont} and \eqref{eq:diff}, the algebraic expression for $\big \langle L^2\big \rangle_{\Omega_0}$ reads as\cite{Tera02}
\begin{align}
  \label{eq:l_full}
  \frac{\big\langle L^2\big\rangle_{\Omega_0}}{l_c^2} &= 2\bigg(\frac{l_p}{l_c}\bigg)^2 \times \Bigg \{ \frac{l_c}{l_p} -1 + \exp \bigg (-\frac{l_c}{l_p} \bigg) \Bigg \} \\ 
  \label{eq:l_part}
  &= 1 - \frac{l_c}{3l_p} + \mathcal{O}\Bigg\{\bigg(\frac{l_c}{l_p}\bigg)^2\Bigg\}.
\end{align}
Let us now assimilate the long axis $\mathbf{u}$ of a stiff chain to the contour-averaged local tangent vector $\mathbf{t}(s)$, so that
\begin{equation}
  \cos \theta_m(0) \equiv \mathbf{t}(0) \cdot \mathbf{u} \cong \mathbf{t}(0) \cdot \big\langle \mathbf{t}\big\rangle_c.
\end{equation}
It follows from the usual definition of the persistence length that
\begin{align*}
  \big\langle\cos \theta_m(0)\big\rangle_{\Omega_0} & = \frac{1}{l_c} \int_0^{l_c} ds \times \big\langle  \mathbf{t}(0)\cdot \mathbf{t}(s) \big\rangle_{\Omega_0} \\
                                                                & =  \frac{1}{l_c} \int_0^{l_c} ds \times \exp  \bigg (-\frac{s}{l_p} \bigg) \\
                                                                & = 1- \frac{\big \langle \theta_m(0)^2\big \rangle_{\Omega_0}}{2} + \mathcal{O}\Big\{\big\langle \theta_m(0)^4\big\rangle_{\Omega_0}\Big\}.
\end{align*}
Thus, for $\theta_m(0) \ll 1$,
\begin{align}
   \frac{ \big \langle\cos^2\theta_m(0)\big \rangle_{\Omega_0}-1}{2} &= - \frac{\big \langle \theta_m(0)^2\big \rangle_{\Omega_0}}{2} \nonumber \\ 
   \label{eq:s_full}
   &= \frac{l_p}{l_c} \bigg\{1-\exp \bigg (-\frac{l_c}{l_p} \bigg)\bigg\} -1 \\
    \label{eq:s_part}
   &= -\frac{l_c}{2l_p} + \mathcal{O}\Bigg\{\bigg(\frac{l_c}{l_p}\bigg)^2\Bigg\}.
\end{align}
Plugging Eqs.~\eqref{eq:l_part} and~\eqref{eq:s_part} into Eq.~\eqref{eq:s_eq} finally yields a simple scaling law for the intra-molecular order parameter of stiff persistent chains,
\begin{equation}
  \label{eq:s_app}
  S_{\Omega_0} \simeq 1-\frac{5l_c}{6l_p}.
\end{equation}

\begin{figure}[h]
  \includegraphics[width=\columnwidth]{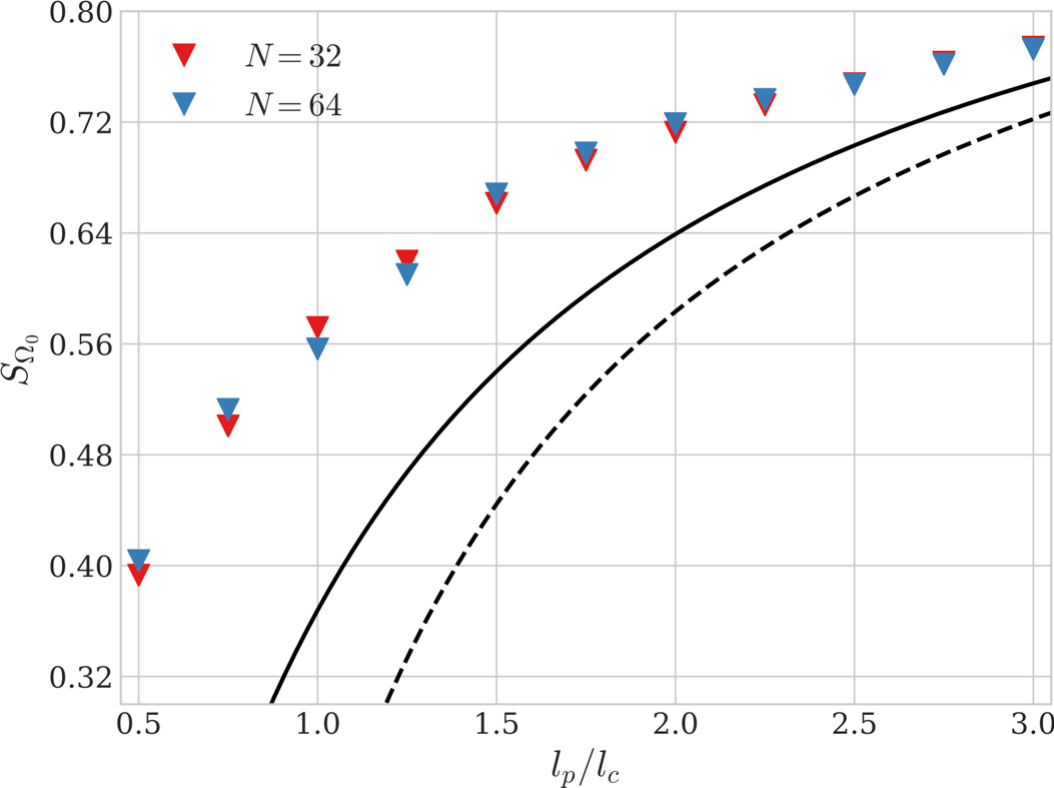}
  \caption{\label{figB1}Intra-molecular bond order parameter $S_{\Omega_0}$ as a function of persistence-to-contour-length ratio for unconfined semi-flexible bead-spring chains. Markers represent the results of single-chain MC simulations with $N=32$ and $N=64$ monomers following the procedure of Sec.~\ref{sec:nematic}. Solid lines denote the theoretical values obtained by plugging the full Eqs.~\eqref{eq:l_full} and~\eqref{eq:s_full} into Eq.~\eqref{eq:s_eq}, and dashed lines those of the truncated asymptotic expansion Eq.~\eqref{eq:s_app}.}
\end{figure}

\end{document}